\documentclass[acmsmall]{acmart}
\usepackage{algorithmic}
\usepackage{graphicx}
\usepackage{textcomp}
\usepackage{xcolor}
\usepackage{caption}
\usepackage{subcaption}
\usepackage{dirtytalk}
\usepackage{multicol}
\usepackage{multirow}
\usepackage[normalem]{ulem}
\usepackage{lipsum} 
\usepackage{soul}
\usepackage{epstopdf}
\usepackage{graphics}
\usepackage{balance}
\usepackage{arydshln}
\usepackage{xspace}
\usepackage{enumitem}
\usepackage{graphicx}
\usepackage{colortbl}
\usepackage{hhline}
\usepackage{ragged2e}
\usepackage{array}
\usepackage{textgreek}

\usepackage[acronyms,nonumberlist,nopostdot,nomain,nogroupskip]{glossaries}
\glsdisablehyper

\newglossary[algh]{hidden}{acrh}{acnh}{Hidden Acronyms}
\newacronym{3gpp}{3GPP}{3rd Generation Partnership Project}
\newacronym{4g}{4G}{4th generation}
\newacronym{5g}{5G}{5th generation}
\newacronym{6g}{6G}{6th generation}
\newacronym{5gc}{5GC}{5G Core}
\newacronym{adc}{ADC}{Analog to Digital Converter}
\newacronym{aerpaw}{AERPAW}{Aerial Experimentation and Research Platform for Advanced Wireless}
\newacronym{ah}{AH}{Authentication Header}
\newacronym{ai}{AI}{Artificial Intelligence}
\newacronym{aimd}{AIMD}{Additive Increase Multiplicative Decrease}
\newacronym{am}{AM}{Acknowledged Mode}
\newacronym{amc}{AMC}{Adaptive Modulation and Coding}
\newacronym{amf}{AMF}{Access and Mobility Management Function}
\newacronym{aops}{AOPS}{Adaptive Order Prediction Scheduling}
\newacronym{api}{API}{Application Programming Interface}
\newacronym{apn}{APN}{Access Point Name}
\newacronym{aqm}{AQM}{Active Queue Management}
\newacronym{ausf}{AUSF}{Authentication Server Function}
\newacronym{avc}{AVC}{Advanced Video Coding}
\newacronym{awgn}{AGWN}{Additive White Gaussian Noise}
\newacronym{balia}{BALIA}{Balanced Link Adaptation Algorithm}
\newacronym{bbu}{BBU}{Base Band Unit}
\newacronym{bdp}{BDP}{Bandwidth-Delay Product}
\newacronym{ber}{BER}{Bit Error Rate}
\newacronym{bf}{BF}{Beamforming}
\newacronym{bler}{BLER}{Block Error Rate}
\newacronym{brr}{BRR}{Bayesian Ridge Regressor}
\newacronym{bsr}{BSR}{Buffer Status Report}
\newacronym{bs}{BS}{Base Station}
\newacronym{bpsk}{BPSK}{Binary Phase-shift keying}
\newacronym{bss}{BSS}{Business Support System}
\newacronym{ca}{CA}{Carrier Aggregation}
\newacronym{caas}{CaaS}{Connectivity-as-a-Service}
\newacronym{cb}{CB}{Code Block}
\newacronym{cc}{CC}{Congestion Control}
\newacronym{ccid}{CCID}{Congestion Control ID}
\newacronym{cco}{CC}{Carrier Component}
\newacronym{cdd}{CDD}{Cyclic Delay Diversity}
\newacronym{cdf}{CDF}{Cumulative Distribution Function}
\newacronym{cdn}{CDN}{Content Distribution Network}
\newacronym{cir}{CIR}{Channel Impulse Response}
\newacronym{cn}{CN}{Core Network}
\newacronym{codel}{CoDel}{Controlled Delay Management}
\newacronym{comac}{COMAC}{Converged Multi-Access and Core}
\newacronym{cord}{CORD}{Central Office Re-architected as a Datacenter}
\newacronym{cornet}{CORNET}{COgnitive Radio NETwork}
\newacronym{cosmos}{COSMOS}{Cloud Enhanced Open Software Defined Mobile Wireless Testbed for City-Scale Deployment}
\newacronym{cots}{COTS}{Commercial Off-the-Shelf}
\newacronym{cp}{C-plane}{Control Plane}
\newacronym{cpu}{CPU}{Central Processing Unit}
\newacronym{cqi}{CQI}{Channel Quality Information}
\newacronym{cr}{CR}{Cognitive Radio}
\newacronym{cran}{CRAN}{Cloud \gls{ran}}
\newacronym{crs}{CRS}{Cell Reference Signal}
\newacronym{csi}{CSI}{Channel State Information}
\newacronym{ct}{CT}{Cipher Text}
\newacronym{csirs}{CSI-RS}{Channel State Information - Reference Signal}
\newacronym{cu}{CU}{Central Unit}
\newacronym{d2tcp}{D$^2$TCP}{Deadline-aware Data center TCP}
\newacronym{d3}{D$^3$}{Deadline-Driven Delivery}
\newacronym{dac}{DAC}{Digital to Analog Converter}
\newacronym{dag}{DAG}{Directed Acyclic Graph}
\newacronym{darpa}{DARPA}{Defense Advanced Research Projects Agency}
\newacronym{das}{DAS}{Distributed Antenna System}
\newacronym{dash}{DASH}{Dynamic Adaptive Streaming over HTTP}
\newacronym{dc}{DC}{Dual Connectivity}
\newacronym{dccp}{DCCP}{Datagram Congestion Control Protocol}
\newacronym{dce}{DCE}{Direct Code Execution}
\newacronym{dci}{DCI}{Downlink Control Information}
\newacronym{dcl}{DCL}{Dear Colleague Letter}
\newacronym{dctcp}{DCTCP}{Data Center TCP}
\newacronym{dl}{DL}{Downlink}
\newacronym{dmr}{DMR}{Deadline Miss Ratio}
\newacronym{dmrs}{DMRS}{DeModulation Reference Signal}
\newacronym{drlcc}{DRL-CC}{Deep Reinforcement Learning Congestion Control}
\newacronym{drs}{DRS}{Discovery Reference Signal}
\newacronym{du}{DU}{Distributed Unit}
\newacronym{e2ap}{E2AP}{E2 Application Protocol}
\newacronym{e2e}{E2E}{end-to-end}
\newacronym{ecaas}{ECaaS}{Edge-Cloud-as-a-Service}
\newacronym{ecn}{ECN}{Explicit Congestion Notification}
\newacronym{edf}{EDF}{Earliest Deadline First}
\newacronym{em}{EM}{Electro-Magnetic}
\newacronym{embb}{eMBB}{Enhanced Mobile Broadband}
\newacronym{empower}{EMPOWER}{EMpowering transatlantic PlatfOrms for advanced WirEless Research}
\newacronym{enb}{eNB}{evolved Node Base}
\newacronym{endc}{EN-DC}{E-UTRAN-\gls{nr} \gls{dc}}
\newacronym{epc}{EPC}{Evolved Packet Core}
\newacronym{eps}{EPS}{Evolved Packet System}
\newacronym{es}{ES}{Edge Server}
\newacronym{esp}{ESP}{Encapsulating Security Payload}
\newacronym{etsi}{ETSI}{European Telecommunications Standards Institute}
\newacronym[firstplural=Estimated Times of Arrival (ETAs)]{eta}{ETA}{Estimated Time of Arrival}
\newacronym{eutran}{E-UTRAN}{Evolved Universal Terrestrial Access Network}
\newacronym{faas}{FaaS}{Function-as-a-Service}
\newacronym{fapi}{FAPI}{Functional Application Platform Interface}
\newacronym{fcc}{FCC}{Federal Communications Commission}
\newacronym{fdd}{FDD}{Frequency Division Duplexing}
\newacronym{fdm}{FDM}{Frequency Division Multiplexing}
\newacronym{fdma}{FDMA}{Frequency Division Multiple Access}
\newacronym{fed4fire}{FED4FIRE+}{Federation 4 Future Internet Research and Experimentation Plus}
\newacronym{fir}{FIR}{Finite Impulse Response}
\newacronym{fit}{FIT}{Future \acrlong{iot}}
\newacronym{fpga}{FPGA}{Field Programmable Gate Array}
\newacronym{fr2}{FR2}{Frequency Range 2}
\newacronym{fs}{FS}{Fast Switching}
\newacronym{fscc}{FSCC}{Flow Sharing Congestion Control}
\newacronym{ftp}{FTP}{File Transfer Protocol}
\newacronym{fw}{FW}{Flow Window}
\newacronym{ga128}{Ga}{Golay Sequence type A}
\newacronym{ge}{GE}{Gaussian Elimination}
\newacronym{glfsr}{GLFSR}{Galois Linear Feedback Shift Register}
\newacronym{gnb}{gNB}{Next Generation Node Base}
\newacronym{gold}{Gold}{Gold}
\newacronym{gop}{GOP}{Group of Pictures}
\newacronym{gpr}{GPR}{Gaussian Process Regressor}
\newacronym{gpu}{GPU}{Graphics Processing Unit}
\newacronym{gtp}{GTP}{GPRS Tunneling Protocol}
\newacronym{gtpc}{GTP-C}{GPRS Tunnelling Protocol Control Plane}
\newacronym{gtpu}{GTP-U}{GPRS Tunnelling Protocol User Plane}
\newacronym{gtpv2c}{GTPv2-C}{\gls{gtp} v2 - Control}
\newacronym{gw}{GW}{Gateway}
\newacronym{harq}{HARQ}{Hybrid Automatic Repeat reQuest}
\newacronym{hetnet}{HetNet}{Heterogeneous Network}
\newacronym{hh}{HH}{Hard Handover}
\newacronym{hol}{HOL}{Head-of-Line}
\newacronym{hqf}{HQF}{Highest-quality-first}
\newacronym{hss}{HSS}{Home Subscription Server}
\newacronym{http}{HTTP}{HyperText Transfer Protocol}
\newacronym{ia}{IA}{Initial Access}
\newacronym{iab}{IAB}{Integrated Access and Backhaul}
\newacronym{ic}{IC}{Incident Command}
\newacronym{ietf}{IETF}{Internet Engineering Task Force}
\newacronym{ifw}{IFW}{Interference Free Window}
\newacronym{imsi}{IMSI}{International Mobile Subscriber Identity}
\newacronym{imt}{IMT}{International Mobile Telecommunication}
\newacronym{iot}{IoT}{Internet of Things}
\newacronym{ip}{IP}{Internet Protocol}
\newacronym{ipsec}{IPsec}{Internet Protocol security}
\newacronym{iq}{IQ}{In-phase and Quadrature}
\newacronym{itu}{ITU}{International Telecommunication Union}
\newacronym{kpi}{KPI}{Key Performance Indicator}
\newacronym{kpm}{KPM}{Key Performance Metric}
\newacronym{kvm}{KVM}{Kernel-based Virtual Machine}
\newacronym{lan}{LAN}{Local Area Network}
\newacronym{los}{LOS}{Line-of-Sight}
\newacronym{ls}{LS}{Loosely Synchronised}
\newacronym{lsm}{LSM}{Link-to-System Mapping}
\newacronym{lstm}{LSTM}{Long Short Term Memory}
\newacronym{lte}{LTE}{Long Term Evolution}
\newacronym{lxc}{LXC}{Linux Container}
\newacronym{m2m}{M2M}{Machine to Machine}
\newacronym{mac}{MAC}{Medium Access Control}
\newacronym{macsec}{MACsec}{Media Access Control Security}
\newacronym{manet}{MANET}{Mobile Ad Hoc Network}
\newacronym{mano}{MANO}{Management and Orchestration}
\newacronym{mc}{MC}{Multi-Connectivity}
\newacronym{mcc}{MCC}{Mobile Cloud Computing}
\newacronym{mchem}{MCHEM}{Massive Channel Emulator}
\newacronym{mcs}{MCS}{Modulation and Coding Scheme}
\newacronym{mec}{MEC}{Multi-access Edge Computing}
\newacronym{mec2}{MEC}{Mobile Edge Cloud}
\newacronym{mfc}{MFC}{Mobile Fog Computing}
\newacronym{mi}{MI}{Mutual Information}
\newacronym{mib}{MIB}{Master Information Block}
\newacronym{miesm}{MIESM}{Mutual Information Based Effective SINR}
\newacronym{mimo}{MIMO}{Multiple Input, Multiple Output}
\newacronym{mgen}{MGEN}{Multi-Generator}
\newacronym{ml}{ML}{Machine Learning}
\newacronym{mlr}{MLR}{Maximum-local-rate}
\newacronym[plural=\gls{mme}s,firstplural=Mobility Management Entities (MMEs)]{mme}{MME}{Mobility Management Entity}
\newacronym{mmtc}{mMTC}{Massive Machine-Type Communications}
\newacronym{mmwave}{mmWave}{millimeter wave}
\newacronym{mpdccp}{MP-DCCP}{Multipath Datagram Congestion Control Protocol}
\newacronym{mptcp}{MPTCP}{Multipath TCP}
\newacronym{mr}{MR}{Maximum Rate}
\newacronym{mrdc}{MR-DC}{Multi \gls{rat} \gls{dc}}
\newacronym{mse}{MSE}{Mean Square Error}
\newacronym{mss}{MSS}{Maximum Segment Size}
\newacronym{mt}{MT}{Mobile Termination}
\newacronym{mtd}{MTD}{Machine-Type Device}
\newacronym{mtu}{MTU}{Maximum Transmission Unit}
\newacronym{mumimo}{MU-MIMO}{Multi-user \gls{mimo}}
\newacronym{mvno}{MVNO}{Mobile Virtual Network Operator}
\newacronym{nalu}{NALU}{Network Abstraction Layer Unit}
\newacronym{nas}{NAS}{Network Attached Storage}
\newacronym{nbiot}{NB-IoT}{Narrow Band IoT}
\newacronym{nfv}{NFV}{Network Function Virtualization}
\newacronym{nfvi}{NFVI}{Network Function Virtualization Infrastructure}
\newacronym{nic}{NIC}{Network Interface Card}
\newacronym{nlos}{NLOS}{Non-Line-of-Sight}
\newacronym{now}{NOW}{Non Overlapping Window}
\newacronym{nrdz}{NRDZ}{National Radio Dynamic Zone}
\newacronym{nsf}{NSF}{National Science Foundation}
\newacronym{nsm}{NSM}{Network Service Mesh}
\newacronym[type=hidden]{nr}{NR}{New Radio}
\newacronym{nrf}{NRF}{Network Repository Function}
\newacronym{nsa}{NSA}{Non Stand Alone}
\newacronym{nse}{NSE}{Network Slicing Engine}
\newacronym{nssf}{NSSF}{Network Slice Selection Function}
\newacronym{ntp}{NTP}{Network Time Protocol}
\newacronym{o2i}{O2I}{Outdoor to Indoor}
\newacronym{oai}{OAI}{OpenAirInterface}
\newacronym{oaicn}{OAI-CN}{\gls{oai} \acrlong{cn}}
\newacronym{oairan}{OAI-RAN}{\acrlong{oai} \acrlong{ran}}
\newacronym{oam}{OAM}{Operations, Administration and Maintenance}
\newacronym[plural=\gls{obu}s,firstplural=Onboard Units (OBUs)]{obu}{OBU}{Onboard Unit}
\newacronym{ofdm}{OFDM}{Orthogonal Frequency Division Multiplexing}
\newacronym{olia}{OLIA}{Opportunistic Linked Increase Algorithm}
\newacronym{omec}{OMEC}{Open Mobile Evolved Core}
\newacronym{onap}{ONAP}{Open Network Automation Platform}
\newacronym{onf}{ONF}{Open Networking Foundation}
\newacronym{onos}{ONOS}{Open Networking Operating System}
\newacronym{oom}{OOM}{\gls{onap} Operations Manager}
\newacronym{opnfv}{OPNFV}{Open Platform for \gls{nfv}}
\newacronym[type=hidden]{o-ran}{O-RAN}{Open \gls{ran}}
\newacronym{orbit}{ORBIT}{Open-Access Research Testbed for Next-Generation Wireless Networks}
\newacronym{os}{OS}{Operating System}
\newacronym{osc}{OSC}{O-RAN Software Community}
\newacronym{osm}{OSM}{Open Street Map}
\newacronym{oss}{OSS}{Operations Support System}
\newacronym{pa}{PA}{Position-aware}
\newacronym{pase}{PASE}{Prioritization, Arbitration, and Self-adjusting Endpoints}
\newacronym{pawr}{PAWR}{Platforms for Advanced Wireless Research}
\newacronym{pbch}{PBCH}{Physical Broadcast Channel}
\newacronym{pcef}{PCEF}{Policy and Charging Enforcement Function}
\newacronym{pcfich}{PCFICH}{Physical Control Format Indicator Channel}
\newacronym{pcrf}{PCRF}{Policy and Charging Rules Function}
\newacronym{pdcch}{PDCCH}{Physical Downlink Control Channel}
\newacronym{pdcp}{PDCP}{Packet Data Convergence Protocol}
\newacronym{pdsch}{PDSCH}{Physical Downlink Shared Channel}
\newacronym{pdu}{PDU}{Packet Data Unit}
\newacronym{pdp}{PDP}{Power Delay Profile}
\newacronym{pf}{PF}{Proportional Fair}
\newacronym{pgw}{PGW}{Packet Gateway}
\newacronym{phich}{PHICH}{Physical Hybrid ARQ Indicator Channel}
\newacronym{phy}{PHY}{Physical}
\newacronym{pl}{PL}{Path Loss}
\newacronym{pt}{PT}{Plain Text}
\newacronym{ptp}{PTP}{Precision Time Protocol}
\newacronym{pmch}{PMCH}{Physical Multicast Channel}
\newacronym{pmi}{PMI}{Precoding Matrix Indicators}
\newacronym{powder}{POWDER}{Platform for Open Wireless Data-driven Experimental Research}
\newacronym{ppo}{PPO}{Proximal Policy Optimization}
\newacronym{ppp}{PPP}{Poisson Point Process}
\newacronym{prach}{PRACH}{Physical Random Access Channel}
\newacronym{prb}{PRB}{Physical Resource Block}
\newacronym{psnr}{PSNR}{Peak Signal to Noise Ratio}
\newacronym{pss}{PSS}{Primary Synchronization Signal}
\newacronym{pucch}{PUCCH}{Physical Uplink Control Channel}
\newacronym{pusch}{PUSCH}{Physical Uplink Shared Channel}
\newacronym{qam}{QAM}{Quadrature Amplitude Modulation}
\newacronym{qci}{QCI}{\gls{qos} Class Identifier}
\newacronym{qoe}{QoE}{Quality of Experience}
\newacronym{qos}{QoS}{Quality of Service}
\newacronym{qtgui}{QT-GUI}{QT Graphical User Interface}
\newacronym{quic}{QUIC}{Quick UDP Internet Connections}
\newacronym{rach}{RACH}{Random Access Channel}
\newacronym{ran}{RAN}{Radio Access Network}
\newacronym[firstplural=Radio Access Technologies (RATs)]{rat}{RAT}{Radio Access Technology}
\newacronym{rcn}{RCN}{Research Coordination Network}
\newacronym{rec}{REC}{Radio Edge Cloud}
\newacronym{red}{RED}{Random Early Detection}
\newacronym{renew}{RENEW}{Reconfigurable Eco-system for Next-generation End-to-end Wireless}
\newacronym{rf}{RF}{Radio Frequency}
\newacronym{rfc}{RFC}{Request for Comments}
\newacronym{rfr}{RFR}{Random Forest Regressor}
\newacronym{ric}{RIC}{\gls{ran} Intelligent Controller}
\newacronym{rlc}{RLC}{Radio Link Control}
\newacronym{rlf}{RLF}{Radio Link Failure}
\newacronym{rlnc}{RLNC}{Random Linear Network Coding}
\newacronym{rmse}{RMSE}{Root Mean Squared Error}
\newacronym{rnis}{RNIS}{Radio Network Information Service}
\newacronym{rr}{RR}{Round Robin}
\newacronym{rrc}{RRC}{Radio Resource Control}
\newacronym{rrm}{RRM}{Radio Resource Management}
\newacronym{rru}{RRU}{Remote Radio Unit}
\newacronym{rs}{RS}{Remote Server}
\newacronym{rsrp}{RSRP}{Reference Signal Received Power}
\newacronym{rsrq}{RSRQ}{Reference Signal Received Quality}
\newacronym{rss}{RSS}{Received Signal Strength}
\newacronym{rssi}{RSSI}{Received Signal Strength Indicator}
\newacronym{rsu}{RSU}{Road-Side Unit}
\newacronym{rtt}{RTT}{Round Trip Time}
\newacronym{ru}{RU}{Radio Unit}
\newacronym{rw}{RW}{Receive Window}
\newacronym{rx}{RX}{Receiver}
\newacronym{s1ap}{S1AP}{S1 Application Protocol}
\newacronym{sa}{SA}{Security Association}
\newacronym{sack}{SACK}{Selective Acknowledgment}
\newacronym{sap}{SAP}{Service Access Point}
\newacronym{sc2}{SC2}{Spectrum Collaboration Challenge}
\newacronym{scef}{SCEF}{Service Capability Exposure Function}
\newacronym{sch}{SCH}{Secondary Cell Handover}
\newacronym{scoot}{SCOOT}{Split Cycle Offset Optimization Technique}
\newacronym{sctp}{SCTP}{Stream Control Transmission Protocol}
\newacronym{sdap}{SDAP}{Service Data Adaptation Protocol}
\newacronym{sd}{SD}{Standard Deviation}
\newacronym{sdk}{SDK}{Software Development Kit}
\newacronym{sdm}{SDM}{Space Division Multiplexing}
\newacronym{sdma}{SDMA}{Spatial Division Multiple Access}
\newacronym{sdn}{SDN}{Software-defined Networking}
\newacronym{sdr}{SDR}{Software-defined Radio}
\newacronym{seba}{SEBA}{SDN-Enabled Broadband Access}
\newacronym{sgsn}{SGSN}{Serving GPRS Support Node}
\newacronym{sgw}{SGW}{Service Gateway}
\newacronym{si}{SI}{Study Item}
\newacronym{sib}{SIB}{Secondary Information Block}
\newacronym{sinr}{SINR}{Signal to Interference plus Noise Ratio}
\newacronym{sip}{SIP}{Session Initiation Protocol}
\newacronym{siso}{SISO}{Single Input, Single Output}
\newacronym{sla}{SLA}{Service Level Agreement}
\newacronym{sm}{SM}{Saturation Mode}
\newacronym{smf}{SMF}{Session Management Function}
\newacronym{smo}{SMO}{Service Management and Orchestration}
\newacronym{sms}{SMS}{Short Message Service}
\newacronym{smsgmsc}{SMS-GMSC}{\gls{sms}-Gateway}
\newacronym{snr}{SNR}{Signal-to-Noise-Ratio}
\newacronym{son}{SON}{Self-Organizing Network}
\newacronym{sptcp}{SPTCP}{Single Path TCP}
\newacronym{srb}{SRB}{Service Radio Bearer}
\newacronym{srn}{SRN}{Standard Radio Node}
\newacronym{srs}{SRS}{Sounding Reference Signal}
\newacronym{ss}{SS}{Synchronization Signal}
\newacronym{sss}{SSS}{Secondary Synchronization Signal}
\newacronym{st}{ST}{Spanning Tree}
\newacronym{svc}{SVC}{Scalable Video Coding}
\newacronym{tb}{TB}{Transport Block}
\newacronym{tcp}{TCP}{Transmission Control Protocol}
\newacronym{tdd}{TDD}{Time Division Duplexing}
\newacronym{tdm}{TDM}{Time Division Multiplexing}
\newacronym{tdma}{TDMA}{Time Division Multiple Access}
\newacronym{tfl}{TfL}{Transport for London}
\newacronym{tfrc}{TFRC}{TCP-Friendly Rate Control}
\newacronym{tft}{TFT}{Traffic Flow Template}
\newacronym{tgen}{TGEN}{Traffic Generator}
\newacronym{tip}{TIP}{Telecom Infra Project}
\newacronym{tls}{TLS}{Transport Layer Security}
\newacronym{tm}{TM}{Transparent Mode}
\newacronym{to}{TO}{Telco Operator}
\newacronym{toa}{ToA}{Time of Arrival}
\newacronym{tr}{TR}{Technical Report}
\newacronym{trp}{TRP}{Transmitter Receiver Pair}
\newacronym{ts}{TS}{Technical Specification}
\newacronym{tti}{TTI}{Transmission Time Interval}
\newacronym{ttt}{TTT}{Time-to-Trigger}
\newacronym{tx}{TX}{Transmitter}
\newacronym{uas}{UAS}{Unmanned Aerial System}
\newacronym{uav}{UAV}{Unmanned Aerial Vehicle}
\newacronym{udm}{UDM}{Unified Data Management}
\newacronym{udp}{UDP}{User Datagram Protocol}
\newacronym{udr}{UDR}{Unified Data Repository}
\newacronym{ue}{UE}{User Equipment}
\newacronym{uhd}{UHD}{\gls{usrp} Hardware Driver}
\newacronym{ul}{UL}{Uplink}
\newacronym{um}{UM}{Unacknowledged Mode}
\newacronym{uml}{UML}{Unified Modeling Language}
\newacronym{up}{U-plane}{User Plane}
\newacronym{upa}{UPA}{Uniform Planar Array}
\newacronym{upf}{UPF}{User Plane Function}
\newacronym{urllc}{URLLC}{Ultra Reliable and Low Latency Communication}
\newacronym{usa}{U.S.}{United States}
\newacronym{usim}{USIM}{Universal Subscriber Identity Module}
\newacronym{usrp}{USRP}{Universal Software Radio Peripheral}
\newacronym{utc}{UTC}{Urban Traffic Control}
\newacronym{vim}{VIM}{Virtualization Infrastructure Manager}
\newacronym{vm}{VM}{Virtual Machine}
\newacronym{vnf}{VNF}{Virtual Network Function}
\newacronym{volte}{VoLTE}{Voice over \gls{lte}}
\newacronym{voltha}{VOLTHA}{Virtual OLT HArdware Abstraction}
\newacronym{vr}{VR}{Virtual Reality}
\newacronym{vran}{vRAN}{Virtualized \gls{ran}}
\newacronym{vss}{VSS}{Video Streaming Server}
\newacronym{wbf}{WBF}{Wired Bias Function}
\newacronym{wf}{WF}{Wired-first}
\newacronym{wi}{WI}{Wireless InSite}
\newacronym{wlan}{WLAN}{Wireless Local Area Network}
\newacronym{pnf}{PNF}{Physical Network Function}
\newacronym{drl}{DRL}{Deep Reinforcement Learning}
\newacronym{mtc}{MTC}{Machine-type Communications}
\newacronym{v2x}{V2X}{Vehicle-to-everything}
\newacronym{cast}{\textit{CaST}}{Channel emulation generator and Sounder Toolchain}
\newacronym{arc}{ARC}{Aerial RAN CoLab}
\newacronym{dsp}{DSP}{Digital Signal Processing}
\newacronym{ota}{OTA}{Over-the-Air}
\newacronym{bom}{BoM}{Bill of Materials}
\newacronym{frand}{FRAND}{fair, reasonable, and non-discriminatory}
\newacronym{ipc}{IPC}{Inter-Process Communications}
\newacronym{uci}{UCI}{Uplink Control Indication}
\newacronym{rdma}{RDMA}{Remote Direct Memory Access}
\newacronym{oran}{Open RAN}{Open Radio Access Network}
\newacronym{tsn}{TSN}{Time-Sensitive Networking}
\newacronym{sp}{S-plane}{Synchronization Plane}
\newacronym{cpri}{CPRI}{Common Public Radio Interface}
\newacronym{ecpri}{eCPRI}{enhanced CPRI}
\newacronym{mp}{M-plane}{Management Plane}
\newacronym{fh}{FH}{fronthaul}
\newacronym{prtc}{PRTC}{Primary Reference Time Clocks}
\newacronym{tgm}{T-GM}{Telecom Grand Master}
\newacronym{tlv}{TLV}{Type, Length, Value}
\newacronym{fp}{FP}{False Positive}
\newacronym{fn}{FN}{False Negative}
\newacronym{bmca}{BMCA}{Best Master Clock Algorithm}
\newacronym{plfs}{PLFS}{Physical Layer Frequency Support}

\definecolor{LightGreen}{RGB}{200,240,200}
\newcolumntype{L}[1]{>{\raggedright\let\newline\\\arraybackslash\hspace{0pt}}m{#1}}
\newcolumntype{C}[1]{>{\centering\let\newline\\\arraybackslash\hspace{0pt}}m{#1}}
\newcolumntype{R}[1]{>{\raggedleft\let\newline\\\arraybackslash\hspace{0pt}}m{#1}}

\newcommand{\M}[1]{\footnotesize\texttt{#1}\normalsize\xspace}
\newcommand{\system}{\ensuremath{\mathsf{TIMESAFE}}\xspace}

\definecolor{lightblue}{RGB}{204, 229, 255}

\AtBeginDocument{%
  }

\setcopyright{cc}
\setcctype{by}
\acmJournal{TOPS}
\acmYear{2025} \acmVolume{1} \acmNumber{1} \acmArticle{1} \acmMonth{1} \acmPrice{}\acmDOI{10.1145/3775060}

\acmJournal{TOPS}
\acmVolume{28}
\acmNumber{5}
\acmArticle{0286}
\acmMonth{12}

\begin{document}

\title{\system: \textbf{T}iming \textbf{I}nterruption \textbf{M}onitoring and \textbf{S}ecurity \textbf{A}ssessment for \textbf{F}ronthaul \textbf{E}nvironments}

\author{Joshua Groen}

\email{joshua.groen@westpoint.edu}
\orcid{0000-0001-5905-7202}
\affiliation{%
  \institution{Northeastern University}
  \city{Boston}
  \state{MA}
  \country{USA}
}
\affiliation{%
  \institution{United States Military Academy}
  \city{West Point}
  \state{NY}
  \country{USA}
}
\email{joshua.groen@westpoint.edu}

\author{Simone Di Valerio}
\orcid{0009-0008-5003-7756}
\affiliation{%
  \institution{Sapienza University of Rome}
  \city{Rome}
  \country{Italy}}

\author{Imtiaz Karim}
\orcid{0009-0000-8680-9932}
\affiliation{%
  \institution{Purdue University}
  \city{West Lafayette}
  \state{Indiana}
  \country{USA}}

\author{Davide Villa}
\orcid{0000-0002-4299-3915}
\affiliation{%
 \institution{Northeastern University}
 \city{Boston}
 \state{Massachusetts}
 \country{USA}}

\author{Yiwei Zhang}
\orcid{0000-0003-2188-8865}
\affiliation{%
  \institution{Purdue University}
  \city{West Lafayette}
  \state{Indiana}
  \country{USA}}

\author{Leonardo Bonati}
\orcid{0000-0002-1511-1833}
\affiliation{%
 \institution{Northeastern University}
 \city{Boston}
 \state{Massachusetts}
 \country{USA}}

\author{Michele Polese}
\orcid{0000-0002-9740-134X}
\affiliation{%
 \institution{Northeastern University}
 \city{Boston}
 \state{Massachusetts}
 \country{USA}}

\author{Salvatore D'Oro}
\orcid{0000-0002-7690-0449}
\affiliation{%
 \institution{Northeastern University}
 \city{Boston}
 \state{Massachusetts}
 \country{USA}}

 \author{Tommaso Melodia}
 \orcid{0000-0002-2719-1789}
\affiliation{%
 \institution{Northeastern University}
 \city{Boston}
 \state{Massachusetts}
 \country{USA}}

\author{Elisa Bertino}
\orcid{0000-0002-4029-7051}
\affiliation{%
  \institution{Purdue University}
  \city{West Lafayette}
  \state{Indiana}
  \country{USA}}

 \author{Francesca Cuomo}
 \orcid{0000-0002-9122-7993}
\affiliation{%
  \institution{Sapienza University of Rome}
  \city{Rome}
  \country{Italy}}

\author{Kaushik Chowdhury}
\orcid{0000-0002-3570-2622}
\affiliation{%
  \institution{University of Texas, Austin}
  \city{Austin}
  \state{Texas}
  \country{USA}}

\renewcommand{\shortauthors}{Groen et al.}

\begin{abstract}
5G and beyond cellular systems embrace the disaggregation of \gls{ran} components, exemplified by the evolution of the \gls{fh} connection between cellular baseband and radio unit equipment.  Crucially, synchronization over the \gls{fh} is pivotal for reliable  5G services. 
In recent years, there has been a push to move these links to an Ethernet-based packet network topology, leveraging existing standards and ongoing research for \gls{tsn}. However, \gls{tsn} standards, such as \gls{ptp}, focus on performance with little to no concern for security. 
This increases the exposure of the open \gls{fh} to security risks. 
Attacks targeting synchronization mechanisms pose significant threats, potentially disrupting 5G networks and impairing connectivity.


In this paper, we demonstrate the impact of successful spoofing and replay attacks against \gls{ptp} synchronization. 
We show how a spoofing attack is able to cause a production-ready O-RAN and 5G-compliant private cellular base station to catastrophically fail within 2 seconds of the attack, necessitating manual intervention to restore full network operations. To counter this, we design a \gls{ml}-based monitoring solution capable of detecting various malicious attacks with over 97.5\% accuracy. 
\end{abstract}





\vspace*{-7em}
\begin{center}
    {\scriptsize
    \textit{This is the authors’ version of the work accepted to 
    \textbf{ACM Transactions on Privacy and Security (TOPS)}, 2025. \\    
    The final version will be available at \url{https://doi.org/10.1145/3775060}.}
    }
\end{center}
\vspace{2em}

\maketitle

\section{Introduction}
\glsresetall

\begin{figure}[tb]
    \centering
    \includegraphics[width=.9\linewidth]{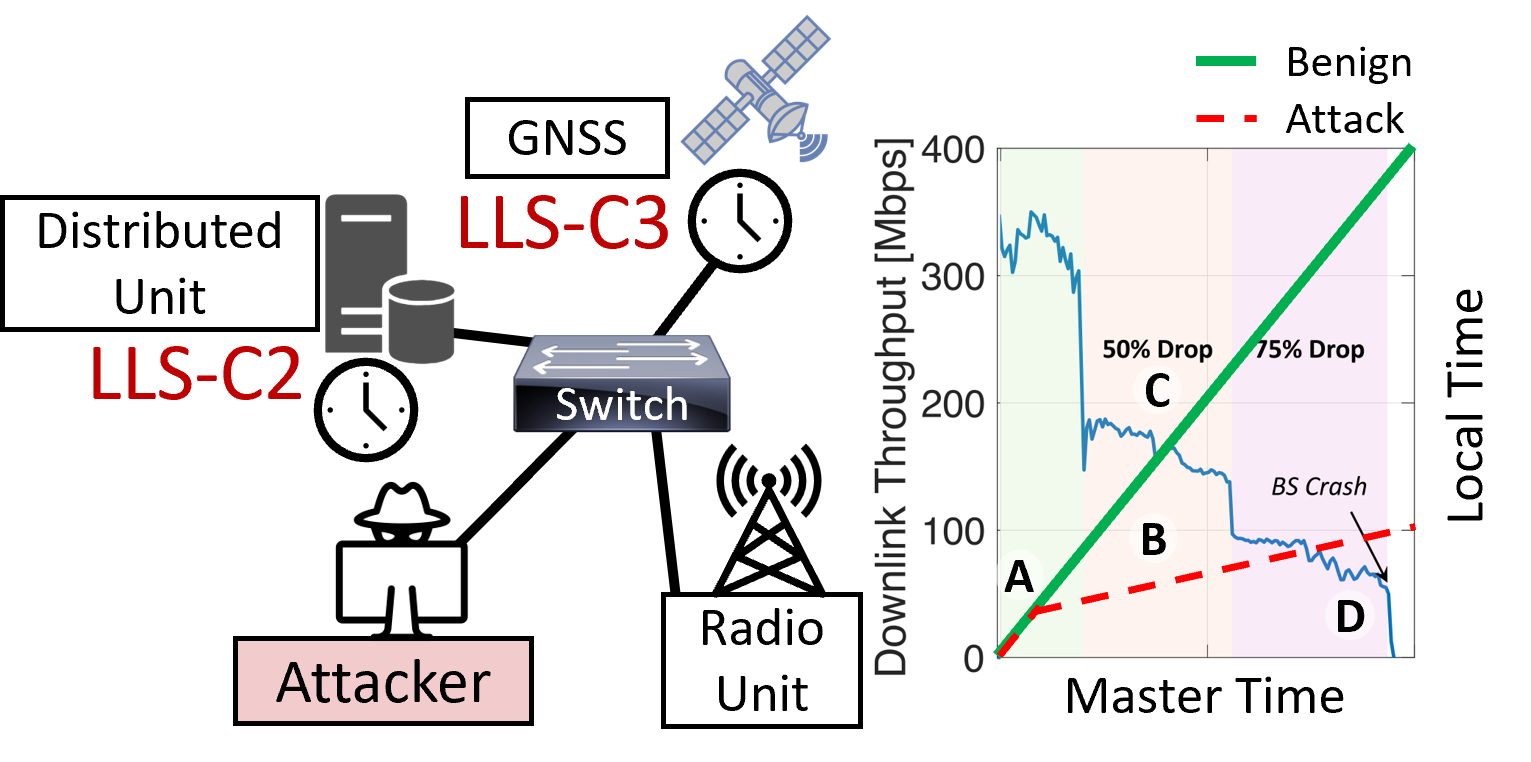}
    \caption{The Open Fronthaul employs PTP to synchronize the master clock with distributed base station components over a switched network. In this scenario, an attacker compromises a device on the network and, at time \textbf{A}, initiates a Spoofing Attack. This leads to gradual synchronization drift (\textbf{B}), first degrading performance (\textbf{C}) and then causing the base station to crash at time \textbf{D}.}
    \Description{Overview of the TIMESAFE system with a graph showing the impact of an attack.}
    \label{fig:overview}
\end{figure}

Recent advancements in 5G and beyond have focused on disaggregating the \gls{gnb} and core networks, leading to the evolution of more open and modular systems. As a result, the \gls{fh} has emerged as a critical component, enabling the separation of radio elements from baseband processing.
Disaggregated FH deployments were originally based on the \gls{cpri}. 
While \gls{cpri} facilitated high-speed, centralized processing of radio signals, it lacked built-in security measures due to its design for single-vendor, co-located deployments~\cite{wong2022security}. The subsequent \gls{ecpri} introduced some security improvements, but significant gaps remain, particularly in synchronization~\cite{ecpri}. The \gls{ecpri}-based O-RAN ALLIANCE's open \gls{fh} now connects disaggregated components from multiple vendors, often using switched Ethernet topologies~\cite{OranWG4,etsiFh}. While Ethernet's flexibility supports diverse traffic types, it also exposes the \gls{fh} to threats, as demonstrated by potential attacks on interconnected devices~\cite{dik2021transport, abdalla2022toward, hung2024security, atalay2023securing, 10.1145/3643833.3656118}. These attacks can disrupt synchronization and cause outages. For example, as shown in Fig.~\ref{fig:overview}, an attacker that gains access to the \gls{fh} network can carry out spoofing attacks that cause synchronization drift, resulting in a reduction in throughput and eventually a complete crash of the base station.
Although there are ongoing efforts by industry, government, and academia to address these security gaps~\cite{wong2022security, ericcson, OranWG11-secreqspec, DoD, abdalla2022toward, groen2024securing}, no comprehensive standards for securing the \gls{fh} have yet been established.

The open \gls{fh} relies on precise timing through the \gls{sp} to perform critical operations like OFDM synchronization, carrier aggregation, and handovers, which require time accuracy within tight margins~\cite{municio2023ran, s_plane_survey}. This synchronization depends on the \gls{ptp}, which, like the \gls{fh}, was initially developed without any security mechanisms~\cite{shi2023ms}. While recent updates to the \gls{ptp} standard include additional security features~\cite[Annex P]{ptp_standard}, these have not been widely implemented in \gls{oran} environments, leaving vulnerabilities that attackers could exploit.

\textbf{Motivation and Scope.} 
Security is critical for the success and widespread adoption of any system, and \gls{oran} has been criticized for lacking robust security mechanisms. In response, the O-RAN ALLIANCE established WG11~\cite{OranWG11-secreqspec} to define a security framework, including zero-trust architectures~\cite{zta}, procedures, and defense mechanisms essential for \gls{oran}'s success~\cite{groen2024securing}. While WG11 acknowledges that attacks on the \gls{sp} can have a high impact, they consider the likelihood low due to the perceived sophistication required by an attacker~\cite{OranWG11-threatmodel}. However, a significant gap remains in analyzing security risks and proposing solutions for the open \gls{fh} \gls{sp} in production-grade \gls{oran} environments. Many researchers lack access to realistic over-the-air systems for testing and validating both attacks and countermeasures. To address this, we developed a Digital Twin system that replicates our production-ready \gls{fh} network environment, enabling rapid experimentation and analysis.

In this paper, we aim to: \emph{(i)} confirm the high impact of \gls{ptp} attacks in a production-ready, 5G and O-RAN-compliant private cellular network; \emph{(ii)} demonstrate that these attacks require relatively low sophistication to execute successfully; \emph{(iii)} provide a Digital Twin framework for further analysis of attacks and development of solutions; and \emph{(iv)} develop and evaluate detection mechanisms, showing that our \gls{ml} model can effectively identify malicious \gls{ptp} attacks with over 97.5\% accuracy across diverse environments.

\textbf{Contributions.} Based on this motivation, we propose \system: \underline{T}iming \underline{I}nterruption \underline{M}onitoring and \underline{S}ecurity \underline{A}ssessment for \underline{F}ronthaul \underline{E}nvironments, a comprehensive framework for assessing the impact and likelihood of attacks against \gls{ptp} in the \gls{fh} of \gls{oran} and 5G networks. \system utilizes machine learning to enhance monitoring and provide highly accurate attack detection. The main contributions of our work include:

\noindent $\bullet$ \textbf{Experimental Analysis:} We conduct the first experimental analysis of the impact of timing attacks in a production-ready private cellular network (see Table~\ref{tab:CompTable}), following O-RAN ALLIANCE procedures~\cite{OranWG11-sectestspec}. Our findings reveal that these attacks can lead to catastrophic outages of the \gls{gnb}, highlighting critical vulnerabilities in \gls{ptp} synchronization mechanisms.

\noindent $\bullet$ \textbf{Attack Demonstration:} We show that timing attacks on \gls{ptp} in the \gls{fh} \gls{sp} are straightforward and require minimal sophistication, countering the belief that such attacks are highly complex.

\noindent $\bullet$ \textbf{Machine Learning Detection:} We develop a state-of-the-art transformer-based \gls{ml} detection mechanism, achieving over 97.5\% accuracy in detecting attacks on a deployed \gls{du}.

\noindent $\bullet$ \textbf{Open Source and Dataset Contributions:} We open source our framework, including our Digital Twin design, the attacks, \gls{ml} models, automation scripts, and a \texttt{.pcap} traffic trace dataset are all open source, to support further research and advancement in O-RAN security~\cite{git_repo}.


\textbf{Responsible Disclosure.} We have disclosed the identified attacks to the O-RAN ALLIANCE Working Group 11 (O-RAN-CVD-002, O-RAN-CVD-003). Our commitment extends beyond this; we are dedicated to collaborating with regulatory organizations and product vendors to enhance security. Through this proactive approach, we aim to ensure the robustness and resilience of the \gls{fh} against future threats.

The rest of the paper is organized as follows. In Section~\ref{background}, we provide background on the \gls{sp} and \gls{ptp}, while Section~\ref{sec:related-work} surveys prior work. Next, we describe the threat model in Section~\ref{threat model} and our test beds in Section~\ref{setup}. We show the impact of the attacks in Section~\ref{a-results}. Then, we demonstrate the need for \gls{ml} based \gls{ptp} monitoring tool in Section~\ref{need ml}, describe our model in \ref{model description} and show state of the art results in Section~\ref{m-results}. Finally, we provide additional insights and directions for research in Section~\ref{discussion}, and draw our conclusions in Section~\ref{conclusion}.

\section{Background}\label{background}


Traditional \gls{ran} architectures are monolithic, leading to vendor lock-in and limited flexibility. The \gls{oran} paradigm addresses these issues with disaggregated, software-based components and open interfaces~\cite{polese2023understanding}, based on the 3GPP 7.2 split between \gls{ru} and \gls{du}. Current O-RAN/ETSI standards~\cite{OranWG4,etsiFh} do not mandate encryption on the \gls{fh} interface due to bandwidth and latency concerns~\cite{groen2024securing}, though some studies suggest using \gls{macsec} for added security~\cite{dik2021transport, dik2023open, cho2021secure}. We start by defining key acronyms in Table~\ref{tab:acronyms}, followed by background on the \gls{sp}, \gls{ptp}, and security.

\begin{table}[htb]
    \centering
    \resizebox{.6\linewidth}{!}{%
        \begin{tabular}{|c|l|}
            \hline
            \rowcolor{lightgray}
            \textbf{Acronym} & \textbf{Definition} \\
            \hline
            \textbf{PTP} & Precision Time Protocol \\
            \textbf{FH} & Fronthaul \\
            \textbf{DU} & Distributed Unit \\
            \textbf{RU} & Radio Unit \\
            \textbf{ML} & Machine Learning \\
            \textbf{O-RAN} & Open Radio Access Network \\
            \textbf{S-plane} & Synchronization Plane \\
            \textbf{UE} & User Equipment (e.g., mobile phone) \\
            \textbf{gNB} & Next Generation Node B (5G base station)  \\
            \textbf{BMCA} & Best Master Clock Algorithm \\
            \textbf{LLS-C} & Lower Layer Split Control Plane\\
            \hline
        \end{tabular}
    } 
    
    \caption{Glossary of commonly used acronyms.}
    \label{tab:acronyms}
\end{table}

\subsection{Fronthaul S-plane}

\begin{figure}[tb]
    \centering
    \includegraphics[width=.9\linewidth]{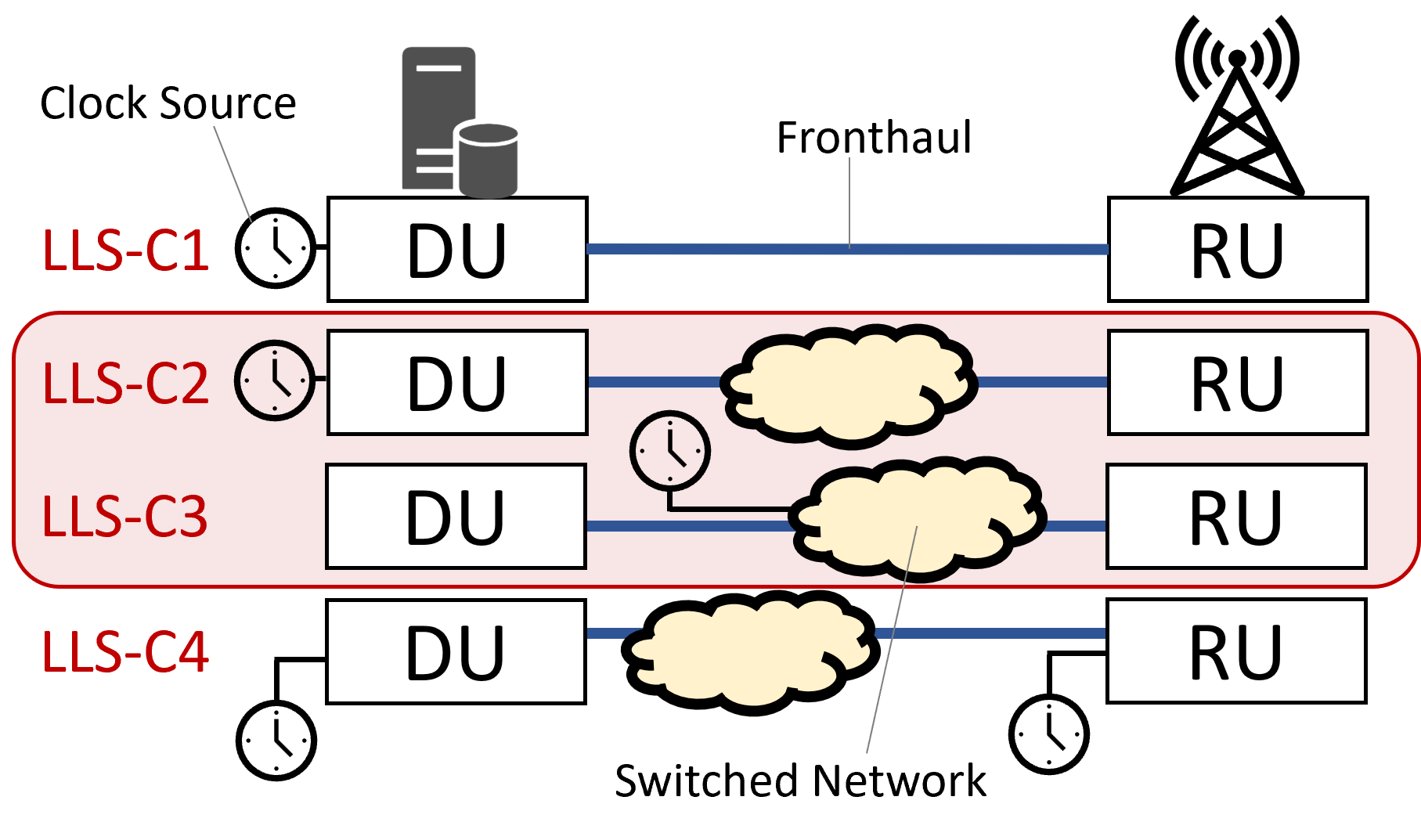}
    \caption{The fronthaul connects the Distributed Unit (DU) to the Radio Unit (RU). There are four Synchronization Plane topologies based on the location of the clock source and switched network. Any topology where the timing information of a single clock has to be synchronized via a packet switch network (e.g. LLS-C2 and LLS-C3) is especially vulnerable to \gls{ptp} attacks because the switched network is used to transport the timing messages. 
    }
    \Description{A figure showing the four configurations for timing in the fronthaul S-plane.}
    \label{fig:sync_overview}
\end{figure}

The \gls{fh} \gls{sp} standards define four clock models and synchronization topologies based on the Lower Layer Split Control Plane (LLS-C): LLS-C1, LLS-C2, LLS-C3, and LLS-C4~\cite{OranWG4}, as illustrated in Fig.~\ref{fig:sync_overview}. Each of these profiles reflects a different method for distributing synchronization between the \gls{du} and \gls{ru}, with varying complexity and exposure to attack.

In LLS-C1, the \gls{du} and \gls{ru} are directly connected, typically via a point-to-point Ethernet link. This straightforward setup uses standard \gls{ptp} signaling between the two endpoints, often with the \gls{du} serving as the master and the \gls{ru} as the slave. With no intermediate devices involved, there are fewer components to misconfigure or compromise, resulting in a relatively low risk of timing manipulation.

LLS-C4 similarly avoids reliance on the transport network by providing each device—both the \gls{du} and one or more \glspl{ru}—with a local GNSS receiver. These receivers discipline the internal clocks using GNSS signals and holdover algorithms, reducing the need for any in-band synchronization messages. This setup removes network-based attack vectors but introduces potential vulnerabilities to GNSS jamming and spoofing, which must be mitigated with resilient oscillator design and fallback mechanisms.

In contrast, LLS-C2 and LLS-C3 rely heavily on the transport network to carry synchronization information, making them far more vulnerable to attack. LLS-C2 places one or more Ethernet switches between the \gls{du} and \gls{ru}, forming a chain in which each switch must function as a Boundary Clock (BC) or Transparent Clock (TC). These intermediate devices forward or adjust \gls{ptp} messages in real time, and their correct behavior is critical for accurate timing. Any misconfiguration, software bug, or compromise of these switches can introduce subtle timing errors or deliberate tampering—both of which may be difficult to detect. LLS-C3 is even more exposed: the \gls{du} and \gls{ru} both receive timing from a central Grandmaster, such as a \gls{prtc}, via a shared switched network. Because neither endpoint provides timing to the other, the entire synchronization path runs through multiple switches and routers, each of which becomes a potential point of failure or compromise. Moreover, both C2 and C3 require careful configuration of PTP parameters, including delay mechanisms, priority levels, and clock class settings, and they depend on hardware timestamping and accurate Best Master Clock (BMC) selection algorithms to maintain time alignment. This increased complexity not only makes these profiles harder to secure and debug but also expands the attack surface significantly. In practice, any adversary capable of accessing the synchronization path—whether through compromised network hardware or malicious software—can degrade or hijack timing across the system, impacting radio coordination, scheduling, and service integrity~\cite{groen2023implementing,OranWG11-threatmodel}.

\subsection{PTP Overview}\label{ptp-overview}


The \gls{ptp} standard (IEEE 1588~\cite{ptp_standard}) ensures precise network clock synchronization in the nanosecond range. It achieves this through hardware time stamping, which minimizes delays from the networking stack~\cite{rezabek2022ptp}. The process involves electing a master clock that synchronizes all slave clocks in the network through a series of exchanges and messages.

\noindent $\bullet$ \textbf{\gls{bmca}.} The initial step in \gls{ptp} involves leader election through the \gls{bmca}. During this process, each clock periodically broadcasts \M{Announce} messages that contain up to nine attributes, which help determine the best clock for master selection. Key attributes include:

\noindent - \textit{Priority1:} A configurable priority setting used as the first comparison feature in the selection process. 

\noindent - \textit{ClockClass:} Indicates the clock's traceability and suitability as a time source. 

\noindent - \textit{ClockAccuracy:} Reflects the clock’s precision.

\noindent - \textit{ClockIdentity:} A unique identifier for the clock, used to resolve ties.

Each clock evaluates the \M{Announce} messages based on the available attributes, selecting the clock with the highest quality as the master. This process is dynamic, allowing for continuous re-evaluation and re-selection if higher-quality clocks are introduced or current attributes change.
    
\noindent $\bullet$  \textbf{Time Synchronization Process.} After being elected, the master clock begins the synchronization process with the slave clocks. 
\textit{Frequency}, \textit{Phase} and \textit{Delay} are critical aspects that influence synchronization.

\textit{Frequency} refers to the rate at which a clock oscillates. To adjust the frequency, the master clock sends a \M{Sync} message and takes a timestamp (\( T_{1} \)). There are two types of clocks: \textit{OneStep} clocks, which insert the timestamp (\( T_{1} \)) directly into the \M{Sync} message, and \textit{TwoStep} clocks, which send a \M{FollowUp} message containing the timestamp (\( T_{1} \)). In the \textit{TwoStep} case, the initial \M{Sync} message has an estimated timestamp or a value of 0. When the slave node receives the \M{Sync} message, it takes a timestamp (\( T_{2} \)). No timestamp is taken when the \M{FollowUp} message is received. 

\begin{figure}[tb]
    \centering
    \includegraphics[width=.9\linewidth]{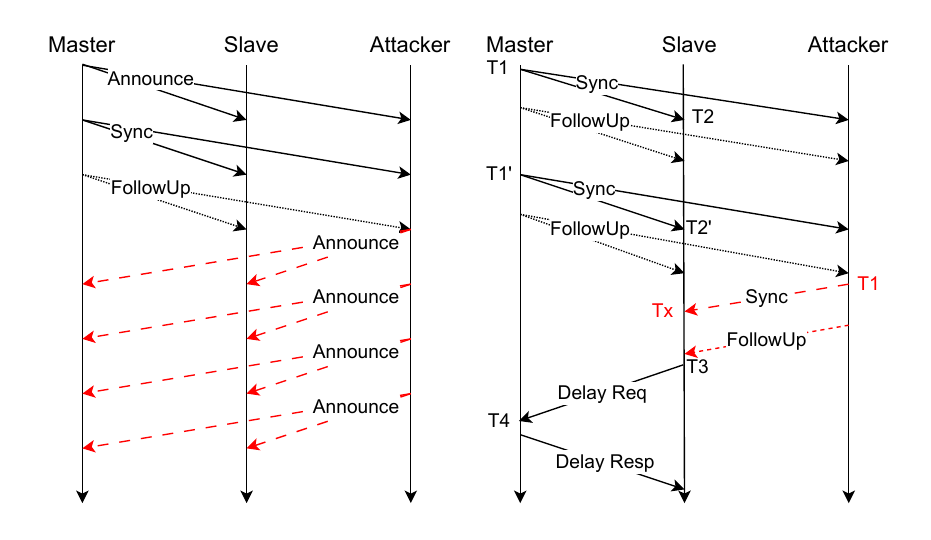}
    \caption{
    The black (solid/dotted) lines represent normal traffic, while the red (dashed) on the left side illustrates the flow of messages during a Spoofing Attack, where an attacker manipulates the \gls{bmca} to become the master. The red (dashed) on the right side illustrates the flow of messages during a Replay Attack,  showing the re-transmission of Sync and FollowUp messages.}
    \Description{A flowgraph demonstrating PTP and attacks.}
    \label{fig:attacks}
\end{figure}

This process is repeated periodically and new timestamps (\( T_{1}' \) and \( T_{2}' \)) are taken again as shown in Fig.~\ref{fig:attacks}. After receiving two \M{Sync} messages, the Slave can calculate the frequency difference relative to the Master clock, which is called $\textit{Drift} = \frac{(T_{2}' - T_{2}) - (T_{1}' - T_{1})}{T_{1}' - T_{1}}$.
By calculating this drift, the Slave can adjust its own frequency to align with the Master clock, ensuring precise synchronization. To meet all \gls{fh} timing requirements, additional \gls{plfs} support, such as SyncE, is required. 
See Appendix (Sec.~\ref{sec:plfs}) for more background on \gls{plfs}.

\textit{Delay} refers to the time it takes for a message or signal to travel from one clock to another through the network. In general, delay can be computed via two mechanisms: (i) the \textit{E2E} mechanism which measures the delay from the master to the slave; and (ii) the \textit{P2P} mechanism which measures the delay between any two nodes irrespective of their role. E2E does not require intermediate nodes to be \gls{ptp} aware, while P2P requires all intermediate switches to be \gls{ptp} aware. In our experiments, we use the E2E mechanism, which is the default configuration. In this case, the master sends a \M{Sync} (and optional \M{FollowUp}) message and takes the timestamp (\( T_{1} \)). The slave receives the \M{Sync} message, takes the timestamp (\( T_{2} \)), sends a \M{DelayReq} and takes the timestamp (\( T_{3} \)). The master receives the \M{DelayReq} message and takes the timestamp (\( T_{4} \)) and stores it. Lastly, the master sends a \M{DelayResp} message with the timestamp (\( T_{4} \)). After receiving the \M{DelayResp} message from the master, the slave can measure the delay between the two nodes: $\textit{Delay} = \frac{(T_{4} - T_{1}) - (T_{3} - T_{2})}{{2}}$. This process, also shown in Fig.~\ref{fig:attacks}, is repeated at defined intervals.

\textit{Phase} refers to the relative alignment of the waveform or signal between two clocks. In \gls{ptp}, phase synchronization ensures that the phase difference between the master clock and the slave clock is minimized, allowing them to maintain consistent timing. After calculating the delay, the slave can also measure the phase, also called $\textit{Offset} = (T_{2} - T_{1}) - \textit{Delay}$.

\subsection{PTP Security}\label{ss:security}

The original \gls{ptp} standard, first published over twenty years ago, did not include any security mechanisms~\cite{shi2023ms}. Subsequent revisions added a security annex, updated in 2019, detailing a flexible, multipronged security approach~\cite{ptp_standard}.

\noindent $\bullet$  \textbf{Prong A} focuses on authentication through an optional \gls{tlv} field, with key management occurring outside of \gls{ptp}. Adding cryptographic functions increases computational cost and processing latency, requiring synchronization adjustments, particularly in resource-constrained environments like FPGA-based \glspl{ru}.

\noindent $\bullet$  \textbf{Prong B} addresses external security mechanisms, including MACsec (Layer 2) and IPsec (Layer 3). However, O-RAN Alliance specifications~\cite{OranWG4} and ITU-T G.8275.1~\cite{8275.1} generally do not permit IP transport. Using MACsec poses challenges; software implementations create virtual interfaces that disrupt hardware timestamping, reducing accuracy significantly. Although hardware-based MACsec solutions have been proposed~\cite{dik2023macsec,dik2023100}, they increase costs, delay, and complexity, reducing the virtualization central to \gls{oran}.

\noindent $\bullet$  \textbf{Prong C} focuses on redundancy (e.g., multiple grandmasters, alternate paths) to mitigate attacks and enhance system resilience. While beneficial, these methods add cost and complexity, conflicting with the push for increased virtualization and decreased costs in \gls{oran}~\cite{polese2023understanding}.

\noindent $\bullet$  \textbf{Prong D} emphasizes monitoring and management. Monitoring \gls{ptp} performance can help detect potential security attacks~\cite{ptp_standard}. The annex suggests parameters to monitor, but does not specify actions to take upon detecting threats. Given the costs associated with the other prongs, it is logical to implement these security measures only when an attack is detected.

\section{Related Work}
\label{sec:related-work}

\begin{table}[htb]
\centering
\resizebox{0.95\linewidth}{!}{
\begin{tabular}
{|C{0.14\textwidth}|C{0.12\textwidth}|C{0.25\textwidth}|C{0.24\textwidth}|C{0.22\textwidth}|}
\hline
\rowcolor{lightgray}
\textbf{Paper} & \textbf{Domain} & \textbf{Attacks Implemented} & \textbf{Mitigation Implemented} & \textbf{Type of Validation} \\ \hline \hline

\cite{decusatis2019impact}  & Data center   &  Spoofing, Comp. Master Clock    &   None     &  Experimental testbed       \\ \hline

\cite{itkin2017security}  & None       &  Spoofing, Comp. Master Clock     &  Prong A, Modifications to \gls{ptp}      & Experimental testbed        \\ \hline

\cite{shereen2019next}   & None       & None     & Prong A       & Experimental testbed  \\ \hline

\cite{rezabek2022ptp}   & None & None     & Prong A  & Experimental testbed  \\ \hline

\cite{moussa2016detection, moussa2018securing, moussa2019extension}  & Smart Grid   & Spoofing, Replay   & Modifications to \gls{ptp}       & Formal model verification   \\ \hline

\cite{shi2023ms} & IoT       & Spoofing     &  Prong C, Modifications to \gls{ptp}   & IoT testbed    \\ \hline

\cite{finkenzeller2024ptpsec}   &   None     &  Spoofing    &   Prong C, Modifications to \gls{ptp}     & Experimental testbed \\ \hline

\cite{s_plane_survey}   &    O-RAN     &   None   &    None    &  None \\ \hline

\cite{dik2021transport,dik2023100,dik2023macsec,dik2023open}   &     O-RAN    &   None   &   Prong B     &  Experimental testbed \\ \hline

\rowcolor{LightGreen} \system    &    O-RAN     &  Spoofing, Replay    &  Prong D      & Production-ready network \\ \hline
 
\end{tabular}
}
   \caption {Overview of prior work on \gls{ptp} security highlighting the application domain studied, the attacks (see Sec.~\ref{ss:attack methods}) and mitigations implemented, and the framework used for validation. Our work is the first to demonstrate the impact of attacks on an O-RAN production-ready environment as well as the first to implement a Prong D (monitoring and management) solution.}
   \label{tab:CompTable}
\end{table}

There is a growing awareness of the need to secure \gls{ptp}. 
Several attacks are demonstrated in a data center context by DeCusatis \emph{et al.}~\cite{decusatis2019impact}. Itkin and Wool developed 
a detailed analysis of threats against \gls{ptp} and proposed using an efficient elliptic-curve public-key signature for Prong A~\cite{itkin2017security}. Shereen \emph{et al.}~\cite{shereen2019next} 
experimentally evaluated implementing Prong A, finding that software-based authentication adds at least an additional 70$\mu$s of delay. Similarly, Rezabek \emph{et al.}~\cite{rezabek2022ptp} evaluated software-based authentication and concluded that there is visible degradation of clock synchronization for each hop in the network with standard deviations between 118 and 571ns.

There are several works that investigate security threats and propose related defense mechanisms for \gls{ptp} in smart grids. 
Moussa \emph{et al.}~\cite{moussa2016detection} proposed adding a new type of \gls{ptp} clock and modifying the \gls{ptp} slave functionality. Then in subsequent work, Moussa \emph{et al.}~ 
proposed adding further message types to aid in detection and mitigation~\cite{moussa2019extension}, and adding feedback from both slaves and master clocks to a reference entity~\cite{moussa2018securing}.

A method targeting Prong C, where clock and network path redundancy are added, is proposed by Shi \emph{et al.}~\cite{shi2023ms}. Likewise, the approach by Finkenzeller \emph{et al.}~\cite{finkenzeller2024ptpsec} uses redundant, or cyclic paths, to detect and mitigate time delay attacks.

Maamary \emph{et al.} 
give an overview of several threats to the \gls{fh} \gls{sp} and discuss possible countermeasures~\cite{s_plane_survey}, though they do not implement either attacks or countermeasures. \gls{fh} security considerations are further described and MACsec is proposed as a solution by Dik \emph{et al.}~\cite{dik2021transport,dik2023100,dik2023macsec,dik2023open}. While these works are the most similar to ours, they primarily address Prong B and could be implemented in parallel to our work. 

In contrast to the above works, our paper is the first to demonstrate successful attacks against \gls{ptp} used for the \gls{sp} causing the gNB to fail and demonstrating how O-RAN-connected devices can be leveraged as attack vectors. We also are the first to propose a \gls{ml}-based method to monitor \gls{ptp} in accordance with Prong D, successfully discriminating between benign and malicious traffic with over 97.5\% accuracy.

\section{Threat model}\label{threat model}

The attack scenario we consider for both the production-ready network and digital twin environment is shown in Fig.~\ref{fig:overview}. In this setup, a switched network connects a software-based \gls{du}, with robust computational and storage capabilities (e.g., an edge server running an open-source network stack), to a more limited \gls{ru}, such as an FPGA-based device. The complexity of the \gls{oran} architecture, with its many interconnected components, heightens security risks\cite{nisReport, nokia}. Although the fronthaul network is protected by IEEE 802.1X, adversaries may bypass this security in a variety of ways. 

\noindent $\bullet$ \textbf{Increased Remote Access.} With its open and multi-vendor architecture, O-RAN relies on a diverse range of suppliers, developers, and service providers, each with varying levels of access and control over networked components. As a result, the number of personnel who can potentially access sensitive infrastructure grows, increasing the risk of misconfigured or compromised authentication processes~\cite{nokia}. IBM's X-Force Threat Intelligence 2023 report identified misuse of valid credentials (36\%) and plaintext credentials on endpoints (33\%) as common initial access vectors in cloud environments~\cite{ibm2023}. Such vulnerabilities are particularly concerning in O-RAN, where distributed network functions and the involvement of multiple parties create more opportunities for configuration errors and unauthorized access to the \glspl{du} ~\cite{abdalla2022toward, verizon}.

\noindent $\bullet$ \textbf{Increased Physical Access.} The distributed nature of \gls{gnb} edge servers, often located in accessible public spaces such as sidewalks, rooftops, or building basements, makes on-site physical access an increasingly viable entry point for attacks on the \gls{fh}~\cite{xing2024criticality}. Physical proximity to fronthaul infrastructure lowers the barrier for potential attackers compared to centralized cloud data centers. Once on-site, an adversary can bypass 802.1X by inserting a rogue device (e.g., a mini PC) into the network to intercept, modify, or replay traffic over already authenticated connections.

\noindent $\bullet$ \textbf{Dependency and Supply Chain Vulnerabilities.} Software-based network functions in 5G, including those in \gls{oran}, are susceptible to supply chain attacks, particularly with the use of unvetted and globally sourced components~\cite{nokia}. A recent government report highlighted that supply chain risks from untrusted hardware and software vendors are significant threats, as adversarial actors could embed backdoors in either the \gls{du} or \gls{ru} to manipulate fronthaul traffic without being detected by 802.1X~\cite{xing2024criticality}. 
Recent studies of several O-RAN software stacks (including those provided by the open-source projects in the O-RAN Software Community) have uncovered numerous high-risk dependencies, misconfigurations, and weak security practices, such as inadequate access control, lack of encryption, and poor secret management across multiple components~\cite{10.1145/3643833.3656118, hung2024security, atalay2023securing, tiberti2022impact, 10588929}.

The combination of increased remote access to \glspl{du} and physical access to the \gls{fh}, along with dependency and supply chain vulnerabilities, makes gaining access to a device connected to the \gls{fh} network not just plausible but increasingly likely~\cite{nokia}. Once an attacker gains access to a device on the switched network, they can observe, replay, or inject \gls{ptp} broadcast packets sent through the switch.

\subsection{Attack Methods}\label{ss:attack methods}
Here we highlight a few of the possible attacks that can compromise \gls{ptp} synchronization.

\noindent $\bullet$  \textbf{Denial-of-Service (DoS) Attack.} A DoS attack floods the network with a high density of \gls{ptp} messages or other types of traffic, overwhelming \gls{ptp} nodes and preventing them from correctly processing timing messages.

\noindent $\bullet$  \textbf{Spoofing Attacks.} In a spoofing attack, an adversary sends malicious \gls{ptp} messages to deceive slave clocks into accepting fake timing information. This method involves impersonating the master clock by sending fake Announce packets or sending false Sync, FollowUp, DelayReq, or DelayResp messages to disrupt legitimate synchronization among nodes in the network.

\noindent $\bullet$  \textbf{Compromised Master Clock.} A compromised master clock poses a severe threat to the \gls{ptp} service, granting attackers control over the network's synchronization source. Attackers can take control of the master clock by exploiting supply-chain vulnerabilities or gaining unauthorized access, enabling manipulation of timing messages and disruption of network operations.

\noindent $\bullet$  \textbf{Replay Attack.} A replay attack in \gls{ptp} involves capturing and retransmitting timing messages, causing devices to synchronize with outdated or modified timestamps. This scenario can disrupt the accuracy of time synchronization and compromise the integrity of time-sensitive operations.

Within the \system framework, we focus on two specific attacks targeting \gls{ptp} synchronization: \textit{Spoofing Attack} and \textit{Replay Attack}. These attacks are chosen for their direct impact on \gls{ptp} synchronization mechanisms. The spoofing attack manipulates the \gls{bmca} to influence master clock election, affecting network-wide synchronization. The replay attack disrupts time synchronization by altering drift, offset, and delay calculations on slave clocks. Details of these attacks are illustrated in Fig.~\ref{fig:attacks} and discussed in the following sections. Other attack types, such as DoS and compromised master clock attacks, involve vulnerabilities beyond \gls{ptp} itself.

\subsubsection{Spoofing Attack}

The \system framework spoofing attack targets \gls{ptp} nodes by sending counterfeit \M{Announce} messages to manipulate the \gls{bmca} leader election and assume the role of the master clock. Initially, the attacker monitors benign traffic to gather critical information about the protocol configuration and node attributes. With this data, the attacker crafts \M{Announce} packets designed to have superior attributes crucial for master clock selection. The \gls{bmca} relies on up to nine features to determine the master clock, starting with \textit{Priority1} (Sec.~\ref{ptp-overview}). The attacker manipulates this value by setting it to 1. Similarly, the attacker sets the values for \textit{ClockClass} 
to 1. The \textit{ClockIdentity} field is created by inserting priority values between the manufacturer ID and device ID portions of the MAC address. The attacker inserts \texttt{ffff} 
ensuring it has higher priority values than the legitimate Master. Other values are copied from the legitimate master.

These crafted messages are sent periodically to maintain the attacker's role as master. Apart from \M{Announce} messages, no synchronization packets 
are sent, and \M{DelayReq} messages from slaves remain unanswered. By disrupting the synchronization process, all clocks operate independently, leading to a gradual drift. Consequently, the attacker introduces synchronization errors among all nodes in the network, causing a complete outage for the \gls{gnb}, as we will show in Section~\ref{ss:real impact} and illustrated in Fig.~\ref{fig:overview}.

\subsubsection{Replay Attack}
The Replay Attack involves sniffing time synchronization messages (\M{Sync} and \M{FollowUp}) from the Master clock, storing, and retransmitting them after a delay. When the Slave clock receives the retransmitted messages, it uses the outdated timestamp \( T_{1} \) and a new timestamp \( T_{x} \) to compute drift, delay, and offset (see Fig.~\ref{fig:attacks}). The formulas in Sec.~\ref{ptp-overview} become: $\textit{Drift} = \frac{(T_{2}' - T_{x}) - (T_{1}' - T_{1})}{T_{1}' - T_{1}}$, $\textit{Delay} = \frac{(T_{4} - T_{1}) - (T_{3} - T_{x})}{2}$, and $\textit{Offset} = (T_{x} - T_{1}) - \textit{Delay}$. The actual difference between when $T_1$ is generated and received in the malicious \gls{ptp} message significantly exceeds the calculated difference, causing offset to increase by orders of magnitude and severely disrupting network synchronization, as we will detail in Section~\ref{ss:dt results}.



\section{Experimental Test Beds}\label{setup}

For \system, we use two experimental test beds to assess network timing interruption attacks, analyze \gls{ptp} security measures in \gls{oran} networks, and evaluate malicious \gls{ptp} detection systems on the \gls{du}. The first test bed is a production-ready 5G network, while the second is a dedicated \gls{oran} digital twin environment for rapid testing, prototyping, and developing attacks and solutions.


\subsection{Production-Ready Network}

Our first experimental analysis is performed on 
a private 5G network, deployed at Northeastern University~\cite{villa2023open}
comprising 8~NVIDIA \gls{arc} nodes, with dedicated \gls{cn} and \gls{fh} infrastructure. 
We utilize this production-ready platform to capture traffic and evaluate attacks on an O-RAN and 5G-compliant system based on a 3GPP 7.2 split, with a combination of open-source components for the higher layers of the stack, as well as commercial devices for the fronthaul infrastructure and the radios~\cite{OranWG4}. 

%

NVIDIA \gls{arc} combines the open-source project \gls{oai}~\cite{oai} for the higher layers of the protocol stack with NVIDIA Aerial, a physical layer implementation accelerated on \gls{gpu} (NVIDIA A100) and with a 7.2 implementation that combines NVIDIA Mellanox smart \glspl{nic}. In an \gls{arc} server, the \gls{gpu} is coupled with the programmable \gls{nic} (a Mellanox ConnectX-6 Dx) via \gls{rdma}, bypassing the CPU for direct packet transfer from the \gls{nic} to the \gls{gpu}. This architecture enables high-speed \gls{du}-side termination of the Open \gls{fh} interface. 

The \gls{fh} infrastructure uses a Dell S5248F-ON switch and a Qulsar QG-2 as the grandmaster clock, distributing \gls{ptp} and SyncE synchronization to the \gls{du} and \gls{ru} with an O-RAN LLS-C3 configuration. For \gls{ptp} synchronization, we use \textit{ptp4l}, an IEEE 1588 compliant implementation for Linux that offers extensive configuration options and enables monitoring and logging of synchronization. The \gls{ru} is a Foxconn 4T4R unit operating in the $3.7-3.8$ GHz band, with \gls{cots} 5G smartphones from OnePlus (AC Nord 2003) as \glspl{ue}~\cite{oneplus}.

While the \gls{du} codebase is open and potentially extensible to secure the S-Plane, the \gls{ru} incorporates a closed-source FPGA-based termination for the \gls{fh} interface, precluding additional security mechanisms in the \gls{fh} environment from being directly enabled. Thus, we also adopt a trace-based approach similar to~\cite{groen2024securing}, configuring a port of the \gls{fh} switch to mirror the \gls{fh} traffic to a server running packet capture.
Specifically, we capture all traffic traversing the \gls{fh} 
for over 20 minutes under several different traffic loads.

\subsection{Digital Twin Environment}\label{ss:emulation}

To leverage these traces we built a digital twin environment using two desktop computers with Intel i9-13900K CPUs and NVIDIA Mellanox ConnectX-4 Lx \glspl{nic} to represent the \gls{du} and \gls{ru}, while a third desktop computer with an Intel Core i7-10700 CPU serves as the attacker. All machines are connected over a private, 10 Gbps capable switched network. Digital twins provide a highly accurate representation of the network and its devices, making them invaluable for designing and testing ML-based algorithms to detect malicious traffic and ongoing attacks~\cite{10179151}. Once trained, \gls{ml}-based control solutions need to be validated and tested in controlled environments to avoid disruptions in production networks~\cite{polese2024colosseum}. Our setup allows for rapid testing, prototyping, and the development of both attacks and solutions, ensuring robust security measures before deployment in real-world scenarios, meeting the need to test and evaluate any proposed \gls{sp} security measures as discussed in \cite{s_plane_survey}.
We use \texttt{.pcap} files captured on the production-ready Open \gls{fh} to accurately emulate the original \gls{fh} and observe the impact the C and U plane traffic has on the PTP protocol operating in the S-Plane~\cite{groen2024securing}. 

On top of the background traffic, we start \gls{ptp} traffic between the \gls{du} and \gls{ru} with \textit{ptp4l}. This mirrors the setup in our production-ready network, which also uses \textit{ptp4l} for \gls{ptp} synchronization.
We use the LLS-C2 profile in our digital twin environment, where the \gls{du} is the Master Clock and the \gls{ru} is our Slave Clock. 
%
The attacker does not actively participate in the \gls{ptp} protocol. Instead, it is capable of observing, replaying, or inserting \gls{ptp} traffic. We discussed the details of our attacks in Section~\ref{ss:attack methods}.  


    
\section{Impact of the Attacks}\label{a-results}

Implementing security measures incurs performance costs~\cite{groen2023implementing,groen2024securing}, so balancing attack risks with the cost of securing the system is crucial. We use the \system framework to evaluate the impact of attacks on the \gls{sp} within our test environments, following the O-RAN interfaces \gls{sp} specification~\cite{OranWG11-sectestspec}. Our assessment focuses on the \gls{gnb}’s availability. According to NIST's National Vulnerability Database (NVD)\cite{NVD}, impact levels are categorized as none, low, or high. Low impact attacks cause performance degradation or intermittent availability without fully denying service, affecting functionalities like Carrier Aggregation and handovers\cite{municio2023ran}. High impact attacks, however, result in complete outages, denying service to all \glspl{ue}.

\subsection{Production-ready Network Impact}\label{ss:real impact}




We test our attacks in the production-ready private 5G network to assess each attack's impact following the recently published guidelines in the O-RAN ALLIANCE testing specifications~\cite{OranWG11-sectestspec}. 
Based on our threat model in Section~\ref{threat model},
we assume that one of the 8~DUs is the compromised machine.

In our initial tests, we found that switch configurations can provide protection against attacks. Malicious packets are not always received by other nodes, as verified through Wireshark captures on both the malicious machine and other nodes. This protection is tied to the switch port's \gls{ptp} role settings. On the Dell S5248F-ON switch, ports set to master only send \gls{ptp} packets, slave mode only receives, and dynamic mode allows both. Additionally, switch configurations can block duplicate MAC addresses. While a properly configured switch prevented our replay attack, assuming correct configurations is risky, especially when \gls{du} and \gls{ru} vendors lack control over network settings, and the complexity and diversity of equipment and configurations in the \gls{fh} path exacerbate this risk~\cite{verizon, ibm2023, rapid7}.

\begin{figure}[tb]
    \centering
    \includegraphics[width=.9\linewidth]{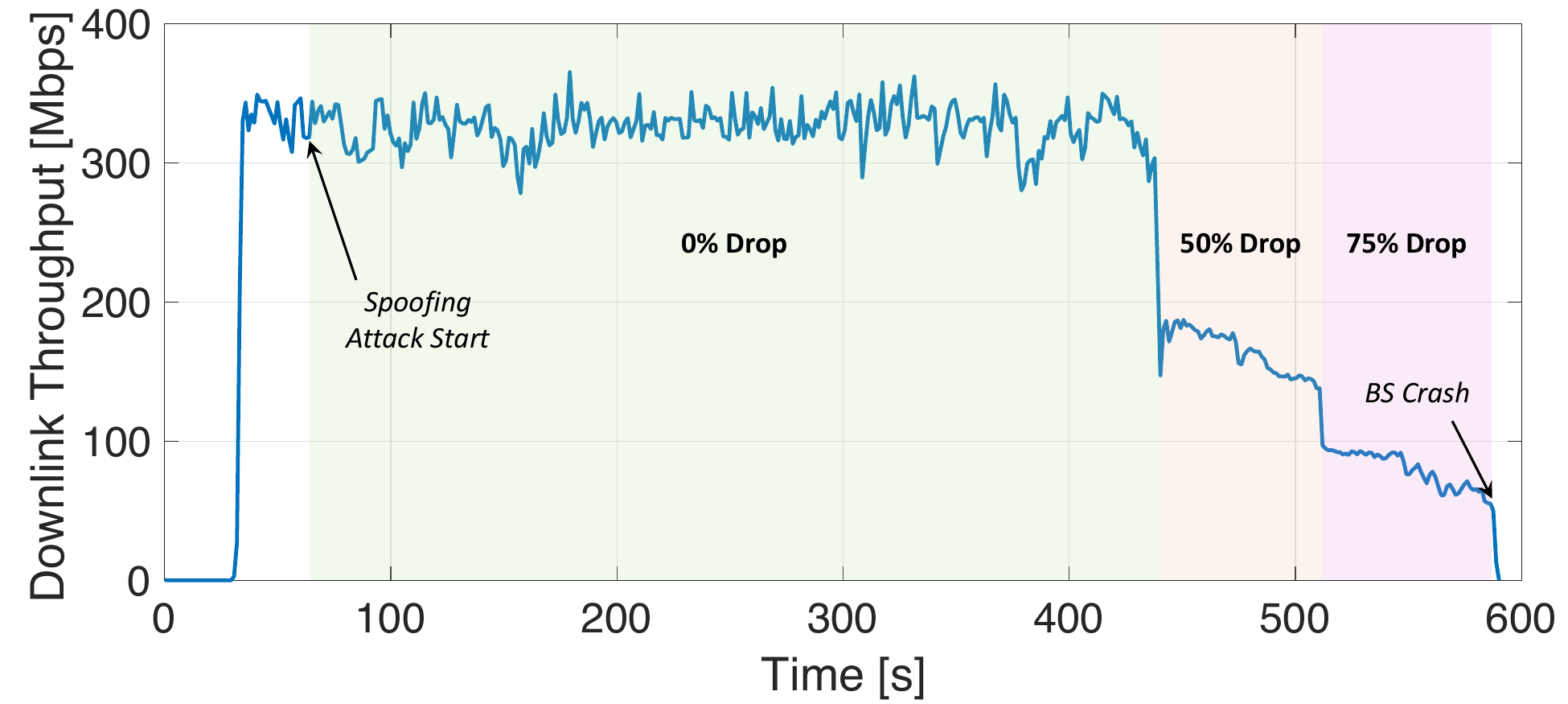}
    \caption{
    Downlink throughput to a \gls{ue} during a spoofing attack with the switch port set to \gls{ptp} master. 
    Approximately 60 seconds into the experiment, the spoofing attack is initiated. Around second 440 there is a 50\% drop in throughput, progressing to a 75\% drop by second 510, and ultimately causing the base station to crash at around second 580. 
    }
    \Description{A graph showing the throughput of an UE during an attack against PTP.}
    \label{fig:SpoofingX5_first_experiment}
\end{figure}

In our first successful attack, we configure the port of our malicious machine with a \gls{ptp} master role. We start our private 5G network, connect the \gls{ue}, and begin sending around 300 Mbps downlink iPerf traffic. After approximately 60 seconds, we launch the spoofing attack, as shown in Fig.~\ref{fig:SpoofingX5_first_experiment}. Monitoring the \gls{ptp} service logs at the \gls{du} reveals that the attack causes the \gls{du} to lose synchronization with the grandmaster clock, leading to a drift in network synchronization as the \gls{du} uses its internal clock. This drift degrades \gls{ue} performance, eventually causing disconnection from the network. About 380 seconds after the beginning of the attack, there is a sudden 50\% drop in throughput. From this point, the throughput continues to drop roughly linearly as the clock drift increases. Finally, the \gls{du} loses synchronization with the \gls{ru}, the throughput drops to 0, and the \gls{ue} disconnects. When this attack stops, the \gls{du} resynchronizes with the grandmaster clock, and the network returns to a healthy state.


\begin{figure}[tb]
    \centering
    \includegraphics[width=.9\linewidth]{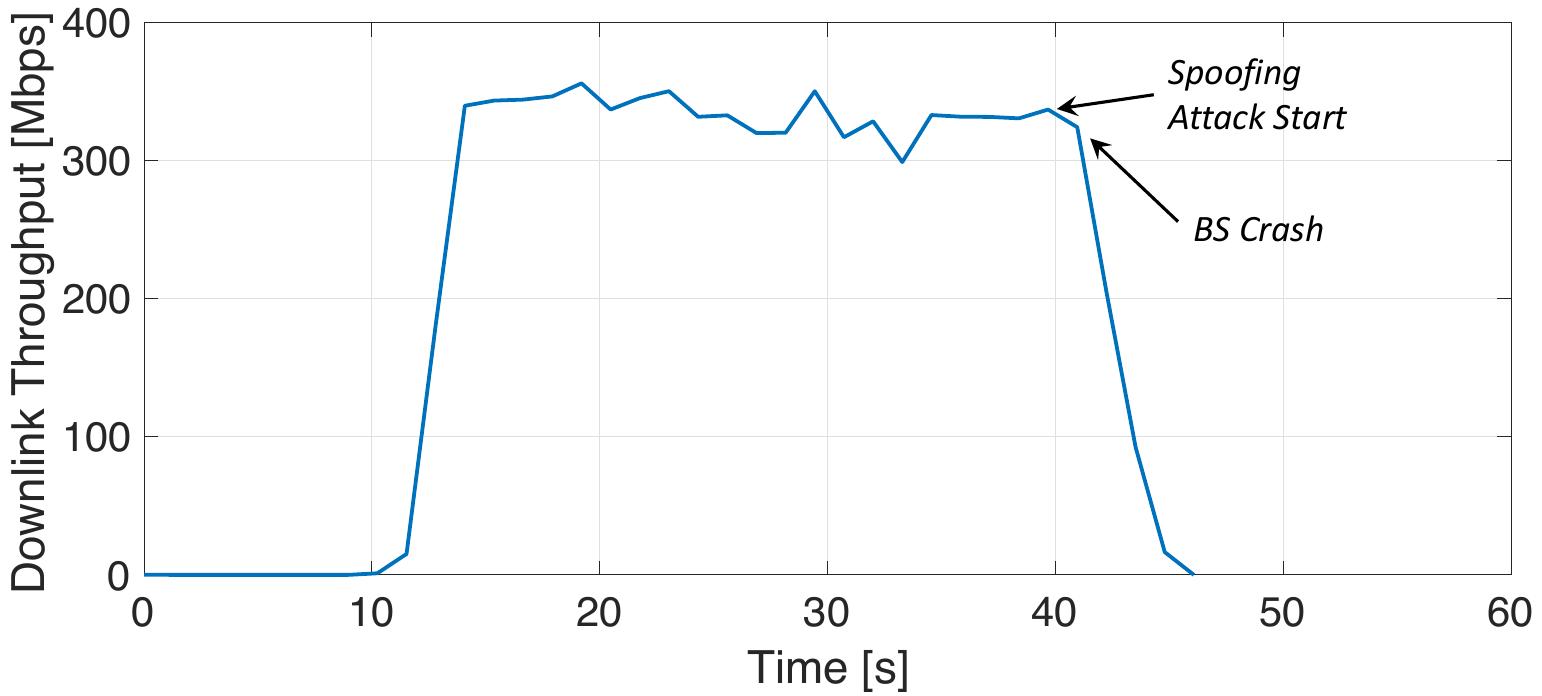}
    \caption{
    Downlink throughput to a \gls{ue} during a spoofing attack with switch ports set to \gls{ptp} dynamic role. Throughput remains stable at around 350 Mbps initially. About 40 seconds in, the spoofing attack begins, causing the 5G cell to drop and the \gls{ue} to lose connectivity. }
    \label{fig:SpoofingX5}
    \Description{A graph showing the throughput of an UE during an attack against PTP.}
\end{figure}


Next, we configure all ports as \gls{ptp} dynamic so that malicious packets can be received by the other nodes and repeat the experiment. This time, the malicious \gls{ptp} packets reach all the machines in the network. Approximately 2 seconds after the attack begins, the \gls{ru} crashes, stopping its operations, causing the 5G cell to drop and the \gls{ue} to lose connection, as shown in Fig.~\ref{fig:SpoofingX5}. While the \gls{du} was able to recover on its own after the first attack, the \gls{ru} required a manual reboot to regain functionality after the second attack. This behavior might be device specific, in our case the Foxconn \gls{ru}. Regardless of any potential vendor-specific variations, these attacks proved to be very powerful, causing significant disruption and necessitating manual intervention to restore full network operations. These tests demonstrate the impact to the \gls{oran} \gls{gnb} is \textit{high}. Failing to secure the \gls{sp} against spoofing attacks results in a complete outage of the \gls{gnb}.




\subsection{Digital Twin Network Impact}\label{ss:dt results}
Next we tested the impact of our attacks on \gls{ptp} functionalities in the Digital Twin environment, focusing on disruptions to synchronization and leader election rather than effects on the \gls{ue}. We employed cyclic attack/recovery patterns with durations of 30/30, 40/20, 50/10 seconds, and continuous attacks for both of the attacks discussed in Sec.~\ref{ss:attack methods}. The attack duration is the period (in seconds) the attack is carried out, while during the recovery time no attack occurs allowing \gls{ptp} synchronization to recover. Each experiment involved four background traffic traces, resulting in 32 tests with durations ranging from 120 to 450 seconds, depending on the traffic. Results from the \gls{ru} confirm significant performance degradation in \gls{ptp}, with varying severity of impact. 

To generate labeled training data, we used logs to identify the timing of attacks and the specific machines involved (based on MAC addresses). A script was written to label each packet in the dataset as either benign (0) or malicious (1) based on whether it came from an attacking machine during the attack window.

Comprehensive results are displayed in the Appendix (Sec.~\ref{sec:addtional results}), but here we highlight the effectiveness of both attacks.



\begin{figure}[tb]
    \centering
    \includegraphics[width=.9\linewidth]{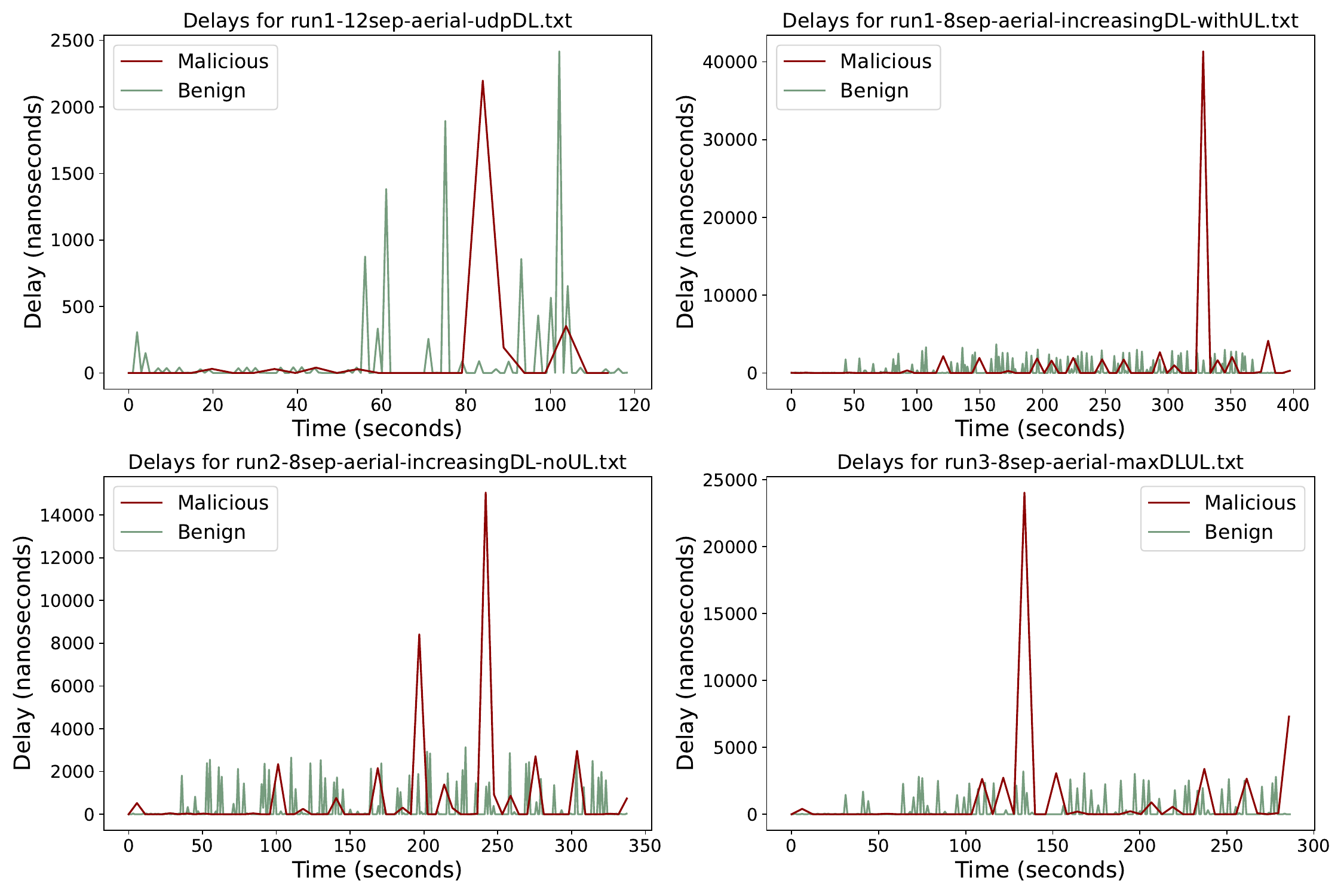}
    \caption{The drift caused by the spoofing attack (dark red) can exceed 40,000~ns, exceeding normal background traffic (light green). During the 50-second spoofing attack period, the \gls{ru} relies on its internal clocks. When the 10-second recovery period begins, a large spike in calculated delay represents the synchronization drift caused by the attack.}
    \Description{A graph of the PTP reported delay during an attack.}
    \label{fig:SpoofingDT}
\end{figure}

\textbf{Spoofing Attack:} Fig.~\ref{fig:SpoofingDT} shows the impact of the 50/10 (50~s attack, 10~s recovery) spoofing attack. Synchronization is disrupted whenever the attacker becomes the master clock, causing the \gls{ru} to rely on its internal clocks, leading to synchronization drift. The impact of this attack depends on the accuracy of each devices internal clock as well as the length of time of the attack. The observable spike in delay is actually reported shortly after the attack ends, when the legitimate master and slave begin synchronizing again. This spike shows the amount of synchronization drift that occurred during the attack. While there is some natural variation in delay due to the background traffic, shown in light green in Fig.~\ref{fig:SpoofingDT}, the announce attack (dark red) shows synchronization drift can exceed 40,000 ns.

\begin{figure}[tb]
    \centering
    \includegraphics[width=.9\linewidth]{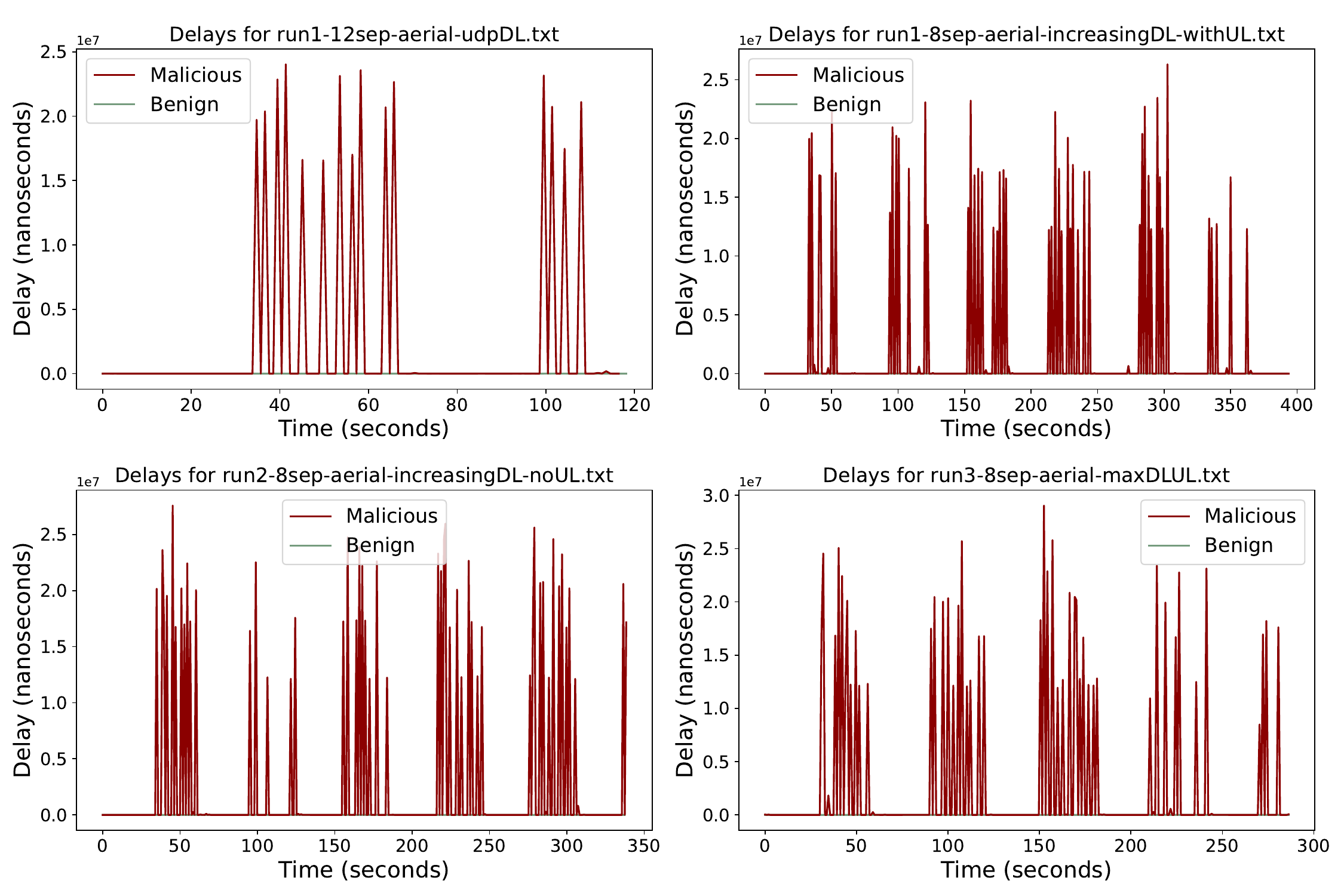}
    \caption{The delay values during the 30-second replay attack (dark red) spike to over 20 million nanoseconds, compared to normal background traffic (light green). The replay attack causes a much higher synchronization drift more quickly than the spoofing attack, and it takes longer for \gls{ptp} to recover. }
    \Description{A graph of the PTP reported delay during an attack.}
    \label{fig:ReplayDT}
\end{figure}

\textbf{Replay Attack:} This attack has a more severe impact on node synchronization, as shown in 
Fig.~\ref{fig:ReplayDT}, where we display the results of the 30/30~s cycle. 
Normally, delay values range between 25 and 45 nanoseconds, but during the Replay attack, these values spike by several orders of magnitude, peaking at over 20 million nanoseconds. Comparing Fig.~\ref{fig:SpoofingDT} and Fig.~\ref{fig:ReplayDT} it is evident that the replay attack is more disruptive than the spoofing attack, causing a high synchronization drift more quickly and requiring more time for \gls{ptp} to fully recover.

\section{Need for ML Based Detection}\label{need ml}

As shown in Section~\ref{a-results}, synchronization attacks can have catastrophic consequences, including complete outages of the \gls{gnb}. However, the virtualized, multi-vendor nature of \gls{oran} and the high precision required for synchronization over the \gls{sp} create challenges for implementing traditional security measures. While hardware-based solutions like MACsec can provide security, they introduce additional costs and complexity to the resource-constrained \gls{ru}~\cite{dik2023macsec,dik2023100}.

To address these limitations, we propose a monitoring system (Prong D) within the \gls{du}, which has more computational resources and flexibility. Given the complexity and variety of potential attacks on synchronization protocols, simple rule-based or threshold mechanisms are insufficient. Instead, \gls{ml} offers a more adaptive approach for several reasons.

\noindent $\bullet$   \textbf{Rule-Based Solutions:} Developing rule-based solutions requires extensive expert knowledge to create specific rules for every potential attack. In Sec.~\ref{ss:heuristic}, we outline the complexity involved in designing expert rules to detect even just two types of attacks.

\noindent $\bullet$  \textbf{Fixed Thresholds:} Rule-based solutions often rely on fixed thresholds, which are challenging to determine optimally. Even with well-designed thresholds, these methods are limited in their ability to adapt to new or evolving threats. In contrast, \gls{ml} models do not require setting fixed thresholds. As we will show in Sec.~\ref{Digital_Twin_Accuracy}, well designed models exhibit resilience againste reshaped attacks. 

\noindent $\bullet$  \textbf{Flexibility:} In contrast to rule-based approaches, a well-trained \gls{ml} model can generalize across various attack types. For instance, we created a new DoS attack by sending a high rate of random or invalid \gls{ptp} messages. Although this attack did not disrupt \gls{ptp} timing, it still indicated malicious activity. Our \gls{ml} model, despite not being specifically trained on this attack, correctly identified over 66\% of the malicious traffic, whereas the heuristic model failed completely. This shows that \system's \gls{ml}-based solution is highly flexible and capable of detecting sophisticated, unpredictable attacks.

\noindent $\bullet$  \textbf{Accuracy.} An effective monitoring solution must deliver accurate results. As we will demonstrate in Sec.~\ref{ss: production ready}, our \gls{ml} model significantly outperforms the rule-based approach, achieving 99\% accuracy in a production environment.

Rule-based systems such as those described in \cite{ptp_standard}
struggle to keep pace with the wide range of attacks, while other heuristic methods require additional clocks or changes to the physical infrastructure \cite{moussa2016detection, shi2023ms, finkenzeller2024ptpsec}. Additionally, legitimate changes in master clocks, changes in network topology, and sequence number resets can often resemble malicious activity, complicating detection efforts. \gls{ml} models excel at identifying patterns and anomalies within large datasets, enabling them to distinguish between benign fluctuations and actual threats. By leveraging \gls{ml}, we can develop robust monitoring systems capable of adapting to new and evolving attack vectors, ensuring reliable network performance and security.

\subsection{Heuristic Detection Results}\label{ss:heuristic}
To highlight the challenges of rule-based systems, We developed a heuristic model to detect attacks using a set of predefined rules.

\textit{Spoofing Attack:} This attack sends numerous \M{Announce} packets to become the master and disrupt synchronization. Our heuristic detects this by monitoring the number of adjacent \M{Announce} packets. Typically, no more than 2 adjacent \M{Announce} messages are seen in normal traffic. If the heuristic detects 4 or more consecutive \M{Announce} packets, it flags an attack.

\textit{Replay Attack:} This sophisticated attack replays benign \M{Sync} and \M{FollowUp} packets with delays, mimicking legitimate traffic. Detection relies on the Sequence ID and Message Type, looking for anomalies like \M{Sync} and \M{FollowUp} messages separated by other packets, Sequence ID values lower than previous packets, and duplicates. The heuristic uses a queue of the last \textit{S} packets to identify these issues, but this method may strain memory resources.

\begin{figure}[tb]
    \centering
    \includegraphics[width=.5\linewidth]{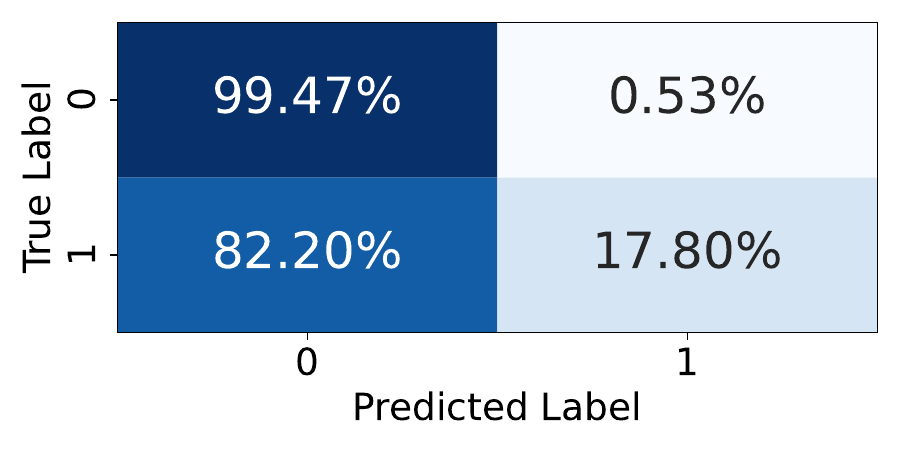}
    \caption{Confusion matrix of the heuristic solution for detection of \gls{ptp} attacks showing an extremely high FN rate.}
    \Description{A confusion matrix.}
    \label{fig:heuristic}
    \end{figure}

Testing under the same conditions we use in Section~\ref{Digital_Twin_Accuracy} showed that while the heuristic effectively detects \M{Announce} attacks, it struggles with overall accuracy. The primary issue is its sensitivity to individual packets: legitimate packets observed during an attack can reset the heuristic, causing it to miss subsequent malicious packets. Results, as shown in Fig.~\ref{fig:heuristic}, indicate poor overall performance due to these limitations.

\section{\system ML-Based Detection}\label{model description}

Traditional heuristic methods are inadequate for detecting all types of malicious packets, given the dynamic and adaptive nature of cyberattacks. Instead, a \gls{ml} approach is better suited to capture the temporal dependencies and patterns in \gls{ptp} data. \gls{ptp} packets, characterized by features such as Ethernet addresses, sequence IDs, message types, and inter-arrival times, form a high-dimensional feature space where subtle anomalies may occur. Unlike rule-based systems, ML models can learn these complex patterns and correlations to identify potential malicious activity.

\subsection{ML Model Selection}
We used our digital twin environment to generate training data, as described in Sec.~\ref{ss:dt results}. 
The data was split into equal-sized chunks (1000 messages each) without overlaps. We randomly selected 80\% of the chunks for training, 10\% for validation, and 10\% for testing. Upon acceptance, we will open-source our complete data set and training pipeline.

We initially evaluated a wide range of \gls{ml} models, as shown in Table~\ref{tab:model_performance}, using the same test traces for all offline accuracy tests. All \gls{ml} models tested had higher accuracy than the rule-based approach in Sec.~\ref{ss:heuristic}. However, accuracy alone is not sufficient; the impact of \glspl{fn} (or missing an attack) is catastrophic, as shown in Section~\ref{ss:real impact}, while the impact of \glspl{fp} is much lower. Therefore, we also examine each model's recall. 

We do recognize that in operational environments, even a modest false positive rate can generate frequent alerts and contribute to operator fatigue, especially when attacks are rare. Heuristic methods may have an advantage of allowing direct adjustment of detection thresholds to accommodate different tolerance for false positives. Although our ML-based approach does not expose simple rule thresholds, the effective sensitivity of the model can be tuned through several mechanisms, such as adjusting the decision threshold on the output probabilities, retraining with modified class weights, or employing cost-sensitive loss functions. These mechanisms allow operators to adapt the false positive/false negative tradeoff to specific deployment needs while maintaining the model’s ability to learn complex, non-linear behaviors that heuristic rules cannot capture. 

Given the clear performance advantage of the CNN and Transformer models, we selected both for further evaluation. The complete confusion matrices for the initial offline results for these models are shown in Fig.~\ref{fig:offline confusion}. Additional details about all models are provided in Appendix~\ref{a:model parameters}.

\begin{table}[htb]
\centering
\resizebox{.7\columnwidth}{!}{ 
    \begin{tabular}{|l|c|c|}
    \hline
    \rowcolor{lightgray}
    \textbf{Model}            & \textbf{Accuracy (\%)} & \textbf{Recall (\%)} \\ \hline
    Heuristic   & 50.95 & 17.80 \\ \hhline{|=|=|=|} 
    Linear Regression         & 75.70                  & 17.90                \\ \hline
    KNN                       & 79.91                  & 51.49                \\ \hline
    Gradient Boosting Classifier & 86.59           & 58.75                \\ \hline
    
    Random Forest             & 84.99                  & 72.00                \\ \hline
    Decision Tree             & 84.26                  & 72.45                \\ \hline
    LSTM                      & 72.97                  & 75.00                \\ \hline
    Naïve Bayes               & 65.72                  & 83.26                \\ \hline
    \rowcolor{lightblue} 
    \textbf{CNN}              & \textbf{97.90}         & \textbf{97.05}       \\ \hline
    \rowcolor{LightGreen} 
    \textbf{Transformer}      & \textbf{97.99}         & \textbf{98.33}       \\ \hline
    \end{tabular}
}
\caption{All machine learning models outperform the heuristic (rule-based) approach in both accuracy and recall on the offline dataset. The CNN and Transformer models achieve significantly higher performance than other methods.}
\label{tab:model_performance}
\end{table}

Our task of classifying packet sequences as malicious or benign parallels sentiment analysis and image classification, where context or spatial features are key. Just as sentiment analysis depends on word context and image classification on visual patterns, our models must capture subtle packet sequence patterns. CNNs excel at extracting spatial features, making them effective at identifying key packet patterns, while transformers are superior in handling sequential data and capturing contextual dependencies. This capability to understand complex patterns intuitively explains why CNN and transformer models outperform others in detecting malicious \gls{ptp} activities.

\begin{figure}
     \centering

     \begin{subfigure}[b]{0.45\textwidth}
        \centering
        \hbox{\hspace{0em} \includegraphics[width=5.5cm]{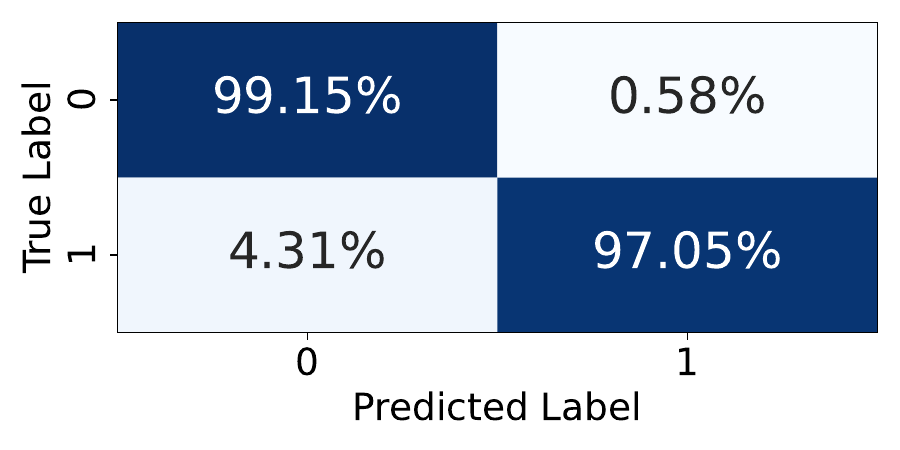}}
        \caption{CNN}
        \label{fig:transformer 3.32}
     \end{subfigure}
     ~
     \begin{subfigure}[b]{0.45\textwidth}
        \centering
        \hbox{\hspace{0em} \includegraphics[width=5.5cm]{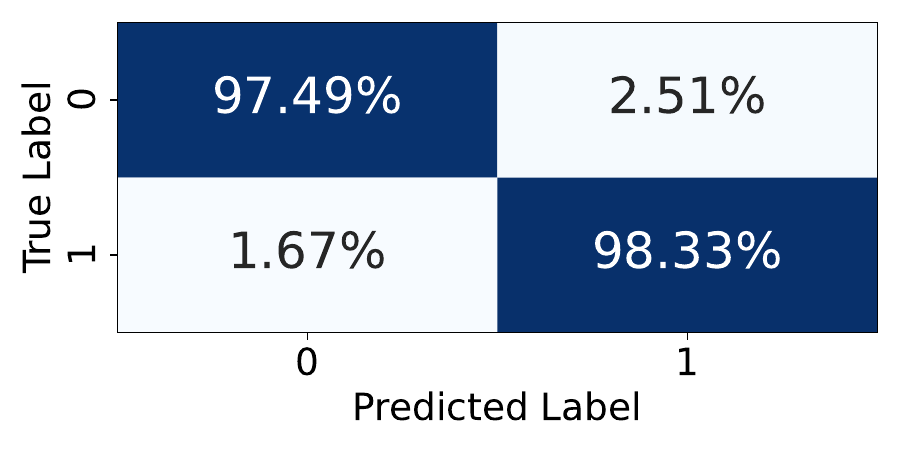}}
        \caption{Transformer}
        \label{fig:transformer 3.40}
     \end{subfigure}
          
    \caption{ Confusion matrices from initial offline testing for the CNN and Transformer. }
    \Description{Confusion matricies.}
    \label{fig:offline confusion}
\end{figure}

\subsection{ML Model Description}

Here we briefly describe the design of the two best performing models, the CNN and transformer. A complete description of each model (see \ref{a:model parameters}) along with code used will be made open source upon acceptance of this article. All the models use the same six input features, $F$: Ethernet source and destination addresses; packet size; \gls{ptp} sequence ID; \gls{ptp} message type; and inter-arrival time. The MAC address helps identify flows and packet direction. The \gls{ptp} message type and sequence ID are crucial for understanding the nature and order of \gls{ptp} messages, aiding in the detection of deviations from expected behavior. Packet length can indicate malformed or malicious packets with extra payload.
Inter-arrival time, specified by the O-RAN ALLIANCE for both announce and sync messages, is another key indicator of potential attacks, though minor variations between packets are expected.

Our CNN model uses a 2D convolutional architecture to process input sequences of fixed length $S$ ($S=32$ for all results presented) and six features ($F=6$), represented as a 1-channel image with dimensions corresponding to the sequence length and feature count. The model has two convolutional layers: the first uses 16 filters with a $3 \times 3$ kernel, followed by ReLU activation and $2 \times 2$ max pooling, reducing spatial dimensions. The second layer applies 32 filters with a $3 \times 3$ kernel, followed by ReLU and max pooling. The output is flattened and passed through two fully connected layers with ReLU and sigmoid activations to produce a binary classification.

\begin{figure}[tb]
    \centering
    \includegraphics[width=.9\linewidth]{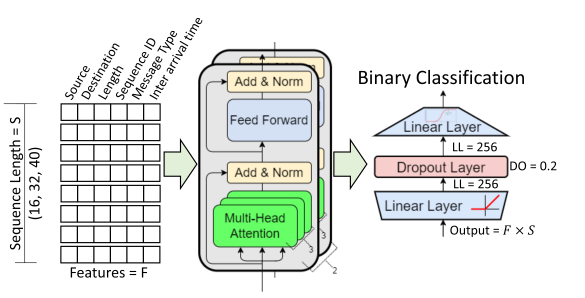}
    \caption{The final \system transformer model uses 2 transformer layers, a fully connected linear layer with ReLU activation function, a dropout layer, and a second linear layer with a sigmoid activation function. The output is a single binary classification for the entire sequence.}
    \Description{A graphic overview of our Transformer architecture.}
    \label{fig:transformer}
\end{figure}

Our transformer-based model (see Fig.~\ref{fig:transformer}) uses only the encoder portion of a full transformer architecture. It processes input sequences with fixed lengths, varied during training and evaluation to include lengths of $S=\{16, 32, 40\}$ packets. These lengths balance the trade-off between incorporating more historical data (and requiring more memory) and improving accuracy and recall. We also treated the number of attention heads $\{2, 3\}$ as a hyper-parameter with two configurations. The transformer encoder outputs a feature matrix of dimension $F \times S$, which is then fed into a fully connected linear layer with a ReLU activation function, followed by a dropout layer with a probability of 0.2 and a final linear layer with a sigmoid activation function. The model produces a single binary classification for the sequence.


Each sequence of length $S$ (used for both CNN and Transformer models) was classified as malicious if any packet in the sequence was labeled as malicious, allowing the models to learn patterns of attack behavior over time. For both the CNN and transformer models, we employed a custom Weighted Binary Cross-Entropy Loss function to address class imbalance and unequal error costs. However, since the initial setting of 1 for both weights yielded satisfactory results, we did not adjust these weights further. In practice, these weights, which correspond to the decision threshold, could be modified to adjust the sensitivity of detection. For example, deployments with more stable network baselines may prefer higher precision to reduce false alarms, whereas high-security environments might prioritize recall to minimize missed attacks. Exploring adaptive threshold tuning or online re-weighting in future work could further improve deployability across diverse scenarios. We used the Adam optimizer with an initial learning rate of 0.001 and configured a StepLR scheduler to reduce the learning rate by a factor of 0.1 every 10 epochs. 
Early stopping was enabled, with monitoring for validation loss improvement and a patience parameter set to 20 epochs. 

Training time depends on both the size of the dataset and the hardware used. However, training is performed offline, making it a one-time cost that does not impact real-time performance. On modern hardware—such as a single NVIDIA A100 GPU on a DGX machine—training never exceeded 30 minutes. This cost is negligible in practice, especially when amortized across many long-term deployments. In the broader context, it is a negligible cost to pay for adding robust security to production environments.

\subsection{Deploying \system}

We deploy \system as a passive detection pipeline to monitor PTP synchronization traffic in the fronthaul segment of the O-RAN architecture. It can run directly on the DU, observing the Ethernet interface connected to the fronthaul switch, or on a separate device connected to a mirror port in the switched network—particularly valuable in LLS-C2 and LLS-C3 profiles where synchronization messages traverse multiple hops and are most vulnerable to attack.

\begin{figure}[tb]
    \centering
    \includegraphics[width=.8\linewidth]{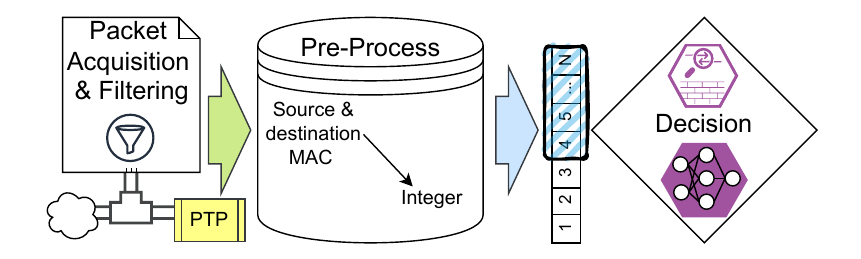}
    \caption{Overview of the \system pipeline deployed for detection in both a production-ready private 5G network and its digital twin environment. The pipeline consists of three modular stages: packet acquisition and filtering, pre-processing, and decision-making.}
    \Description{A graphical overview of our system.}
    \label{fig:pipeline deployment}
\end{figure}

The pipeline consists of three modular stages: packet acquisition and filtering, pre-processing, and decision-making, as shown in Fig.~\ref{fig:pipeline deployment}. The first stage captures Ethernet frames and filters for PTP packets, extracting six features: source and destination MAC addresses, frame length, PTP sequence ID, message type, and inter-arrival time. These are passed via thread-safe queues to the pre-processing stage, which maps MAC addresses to integers to abstract away device-specific identifiers and enable generalizable traffic pattern learning.

The decision-making stage maintains a sliding window of size $S$ and evaluates each sequence using either a machine learning model or a rule-based algorithm. After every two new PTP packets, the window advances, and a new classification decision is made. The model's detection process is designed to operate in real time, with low-latency inference on both the CNN and Transformer models—even on moderate hardware. While the Transformer architecture is computationally heavier due to its attention mechanism, both models sustain real-time decision-making without introducing bottlenecks in both of our testbeds. This approach enables prompt detection of anomalies while keeping latency low. The modular design also allows each stage to be updated independently, supporting flexible deployment and future model improvements.

\section{\system Detection Results}\label{m-results}
This section demonstrates that \system accurately detects attacks in a production-ready 5G network and shows its resilience to reshaping attacks in our Digital Twin environment.

\subsection{Production-ready Detection Accuracy}\label{ss: production ready}

\begin{figure}
     \centering

     \begin{subfigure}[b]{0.45\textwidth}
        \centering
        \hbox{\hspace{0em} \includegraphics[width=5.5cm]{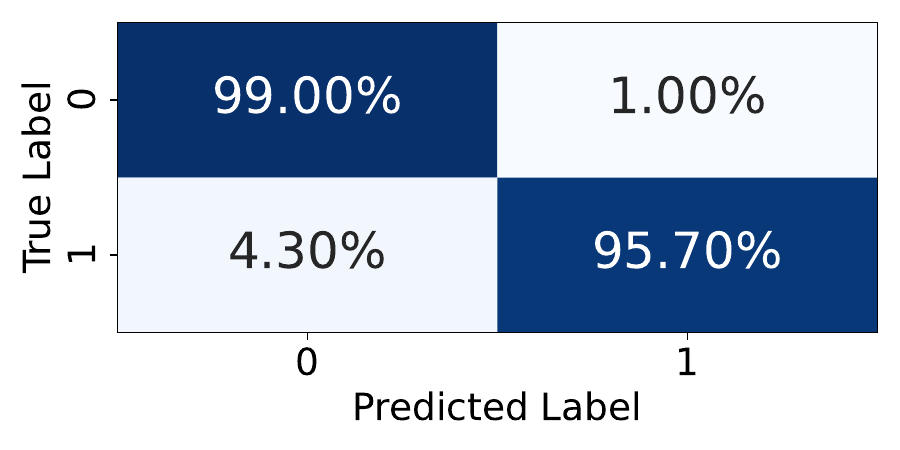}}
        \caption{Transformer}
        \label{fig:transformer prod}
     \end{subfigure}
     ~
     \begin{subfigure}[b]{0.45\textwidth}
        \centering
        \hbox{\hspace{0em} \includegraphics[width=5.5cm]{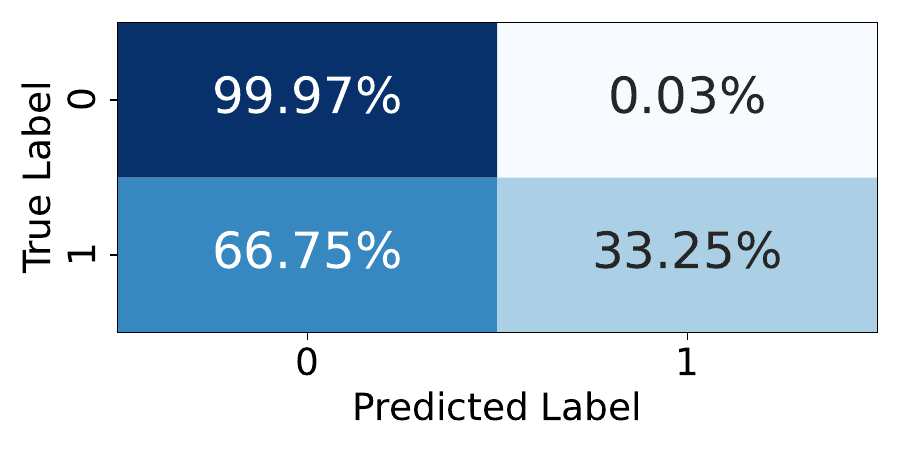}}
        \caption{CNN}
        \label{fig:cnn prod}
     \end{subfigure}
          
    \caption{ Confusion matrices for production environment attack detection. The transformer greatly outperforms the CNN, showing it is able to adapt to new environments. }
    \Description{Confusion matricies.}
    \label{fig:prod attack}
\end{figure}

We evaluated the \system detection pipeline using a trace (\texttt{.pcap}) captured during a successful attack in the production-ready environment, as illustrated in Fig.~\ref{fig:SpoofingX5}. The results demonstrate that the transformer model with 2 attention heads and a slice length of 32 significantly outperformed other transformer configurations, achieving an accuracy exceeding 99\%. The confusion matrix in Fig.~\ref{fig:transformer prod} illustrates the model's strong performance, highlighting its ability to accurately distinguish between malicious and benign traffic in a real-world setting.

In contrast, Fig.~\ref{fig:cnn prod} shows that the performance of the CNN model was substantially degraded when applied to the new environment. This suggests that the CNN struggled to generalize to unseen data and failed to adapt to the subtle contextual variations present in the production environment. The transformer model, however, maintained its robustness, effectively identifying the nuanced patterns and indicators of the attack, even in a more complex and dynamic setting. This reinforces the transformer’s ability to generalize well to new conditions, capturing intricate relationships across sequences that may vary from one environment to another. 

While the CNN exhibited slightly lower false positives in controlled digital twin experiments, its inability to generalize to the production environment led to a significant drop in overall detection reliability. In contrast, the transformer’s small increase in false positives is outweighed by its robustness to environmental shifts and unseen attack patterns. In practical deployments where detection failures are more costly than occasional false alarms, this tradeoff favors the transformer as the more stable and effective choice.

\subsection{Digital Twin Detection Accuracy}\label{Digital_Twin_Accuracy}

To further validate our \gls{ml} model, we deployed \system in a digital twin environment that simulates the production-ready network. This setup provides a realistic testbed for evaluating the model’s performance under a wide variety of conditions. While running full \gls{ptp} with \textit{ptp4l}, we keep system time updates on the \gls{du} and attacker disabled, allowing us to log prediction outcomes with accurate \gls{du} timestamps. The attacker logs the attack type and its timestamps, enabling us to measure decision accuracy. Analysis reveals that we can detect attacks with just one or two malicious \gls{ptp} packets reaching the \gls{du}.

\begin{figure}
     \centering
        \begin{subfigure}[b]{0.45\textwidth}
        \centering
        \hbox{\hspace{0em} \includegraphics[width=5.5cm]{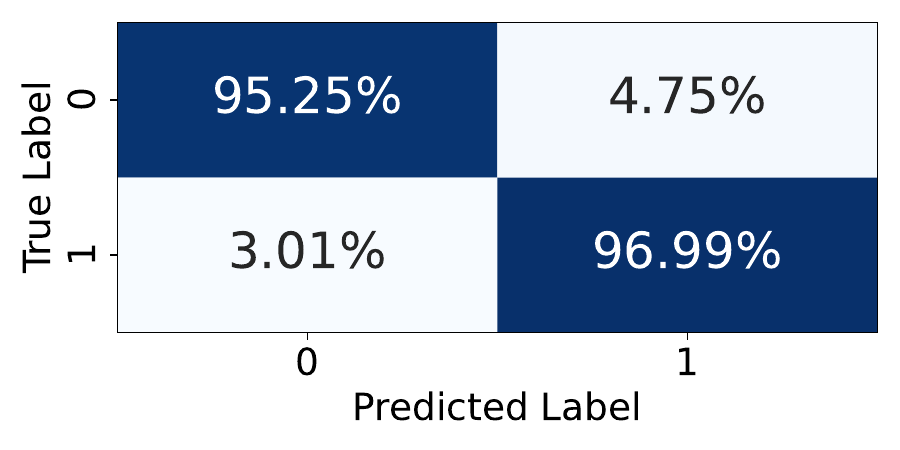}}
        \caption{Transformer}
        \label{fig:overall_confusion_matrix_bg}
     \end{subfigure}
     ~
      \begin{subfigure}[b]{0.45\textwidth}
        \centering
        \hbox{\hspace{0em} \includegraphics[width=5.5cm]{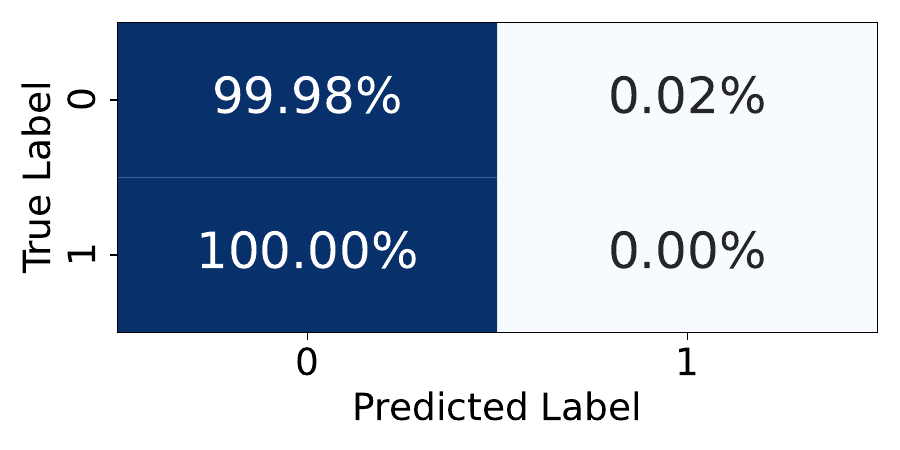}}
        \caption{CNN}
        \label{fig:cnn_dt}
     \end{subfigure}

     \caption{Confusion matrices for evaluating resilience to reshaped attacks with randomized attack and recovery durations. The transformer shows resilience to this strategy change, while the CNN fails to generalize.
     }
     \Description{Confusion matricies.}
     \label{DT test}
\end{figure}

During the experiment, several tests were performed. At the start of each test, the \gls{du}, \gls{ru}, and attacker exchanged initial messages to confirm readiness. Once prepared, they initiated a 5-minute test. Throughout the test, the \gls{du} continuously logged predictions with timestamps, while the attacker recorded the attack types, along with their start and end times, and the duration of both the attack and recovery phases. This cycle repeated for the entire test duration, alternating between attack and recovery phases. Recall that during training, we used cycles of 30/30, 40/20, 50/10 seconds, and continuous attack, as described in Section~\ref{ss:dt results}. To evaluate the models' ability to generalize to altered attack patterns, the attacker in this experiment randomly selected an attack duration from the continuous range $[10, 30]$ seconds and a recovery duration from the continuous range $[40, 60]$ seconds. The detection results from this experiment are shown in Fig.~\ref{DT test}. The transformer demonstrated excellent resilience against the reshaped attack, while the CNN completely failed to detect it. This outcome highlights the transformer's robust ability to adapt to previously unseen attack strategies, confirming greater suitability for real-world deployment where attack patterns can be unpredictable.

    
\section{Discussion}\label{discussion}

The \gls{oran} framework's push towards increased virtualization and softwarization~\cite{afolabi2018network, ko2024edgeric} necessitates security solutions that align with this trend. Prong D (Sec.~\ref{ss:security}), which involves monitoring, is well-suited to the softwarized \gls{du} environment. However, future work should prioritize several additional key areas.

\textbf{Attacks in Other Synchronization Topologies.} Our focus on LLS-C2 and LLS-C3, due to their high vulnerability, does not encompass all risks. For instance, jamming GNSS signals could affect LLS-C4, and supply chain attacks on primary reference clocks pose additional threats. Network redundancy, as suggested in Prong C, enhances protection and resilience, potentially mitigating such threats through redundant paths and clocks~\cite{xing2023enabling, lazarev2023resilient, shi2023ms, finkenzeller2024ptpsec}.

\textbf{Additional Attacks Against \gls{ptp}.} While \system currently addresses two types of attacks, other threats like packet removal and selective delay manipulation also merit investigation~\cite{s_plane_survey}. Expanding \system to include multi-class classification for various attack types could refine threat detection and response strategies.

\textbf{Authentication.} Enhancing \gls{ptp} security through authentication, as outlined in Prong A, is crucial. Implementing digital signatures can verify message integrity and authenticity, preventing replay and spoofing attacks. However, integrating these mechanisms must balance security with the real-time performance needs of \gls{ptp} in \gls{oran}. Efficient, lightweight cryptographic algorithms are essential to avoid introducing latency that disrupts timing synchronization. Additionally, the cost of upgrading \gls{ru} equipment, particularly FPGA-based systems, to support these algorithms must be weighed against the possible risks to enable an informed, risk-based decision on where and how to best implement this type of solution.

Our detection-based approach is designed to complement these preventive measures. Unlike cryptographic or hardware-based defenses, which may require system upgrades or protocol changes, \system operates on traffic patterns alone and can be deployed as a lightweight, software-only monitor in existing networks. While authentication mechanisms aim to prevent certain attacks, \system helps identify when those defenses may be necessary, have failed or when other forms of manipulation (e.g., asymmetric delay) occur. Thus, detection and defense should be viewed as complementary components of a comprehensive security posture for PTP synchronization.

\section{Conclusion}\label{conclusion}

Securing \gls{ptp} in the \gls{sp} within the \gls{oran} framework is essential, as demonstrated by our findings. Successful attacks can lead to severe disruptions, such as complete \gls{gnb} outages requiring manual restoration, underscoring the need for robust security in multi-vendor environments that expand attack surfaces.

Addressing these threats involves balancing security costs against performance and financial implications. While robust measures like encryption introduce latency and computational overhead, \system offers an effective solution. Our machine learning-based monitoring system achieves over 97.5\% accuracy in detecting malicious attacks, providing a cost-efficient approach with minimal additional cost.

By using advanced machine learning techniques, \system enables real-time detection of threats, allowing for targeted application of higher-cost security measures only when necessary. This adaptive strategy optimizes security and performance, highlighting the importance of tailored solutions to safeguard \gls{ptp} synchronization in \gls{oran} networks. The demonstrated vulnerabilities in the \gls{fh} call for proactive, cost-effective security measures to prevent significant network disruptions.

\begin{acks}
This article is based upon work partially supported by the U.S.\ National Science Foundation under grants CNS-1925601 and CNS-2112471 and by SERICS (PE00000014) 5GSec project under the MUR National Recovery and Resilience Plan funded by the European Union - NextGenerationEU.
The opinions in the work are solely of the authors and do not reflect those of the the U.S. Military Academy, U.S. Army, or the Department of Defense.

\end{acks}

\bibliographystyle{ACM-Reference-Format}
\bibliography{references}

\newpage

\appendix

\section{Appendix}
\label{Sec:Appendix}

\subsection{Additional Attack Results}\label{sec:addtional results}
In this section we present additional results demonstrating the impact of our \gls{ptp} attacks in terms of reported delay. The delay during normal conditions with \gls{fh} background traffic is shown in light green, while the delay caused by our attacks is shown in dark red.

\begin{figure}[htb]
    \centering
    \includegraphics[width=.9\linewidth]{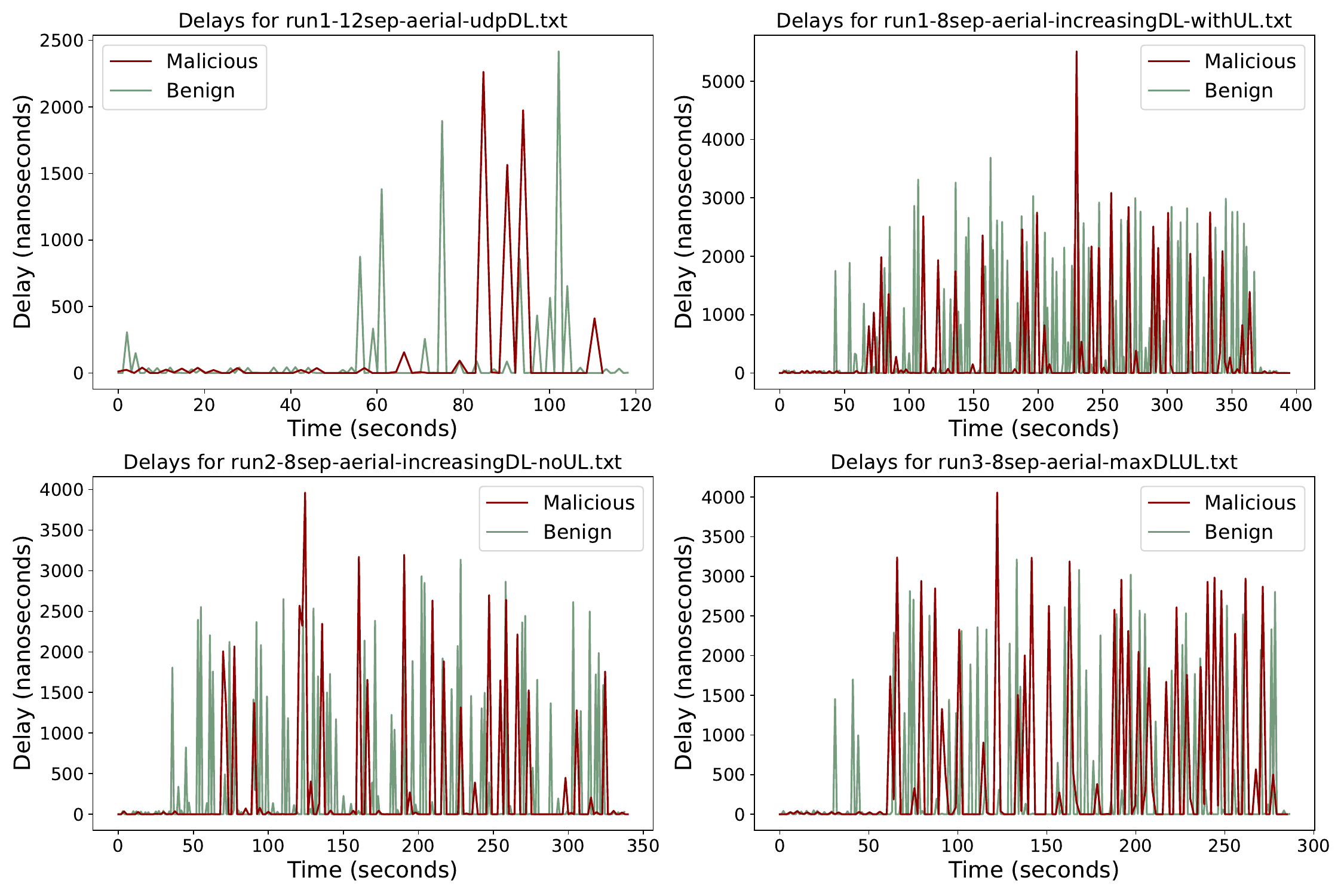}
    \caption{Impact of the 30/30 second spoofing attack. }
    \Description{A graph of the PTP reported delay during an attack.}
    \label{fig:Spoofing30/30}
\end{figure}

\begin{figure}[htb]
    \centering
    \includegraphics[width=.9\linewidth]{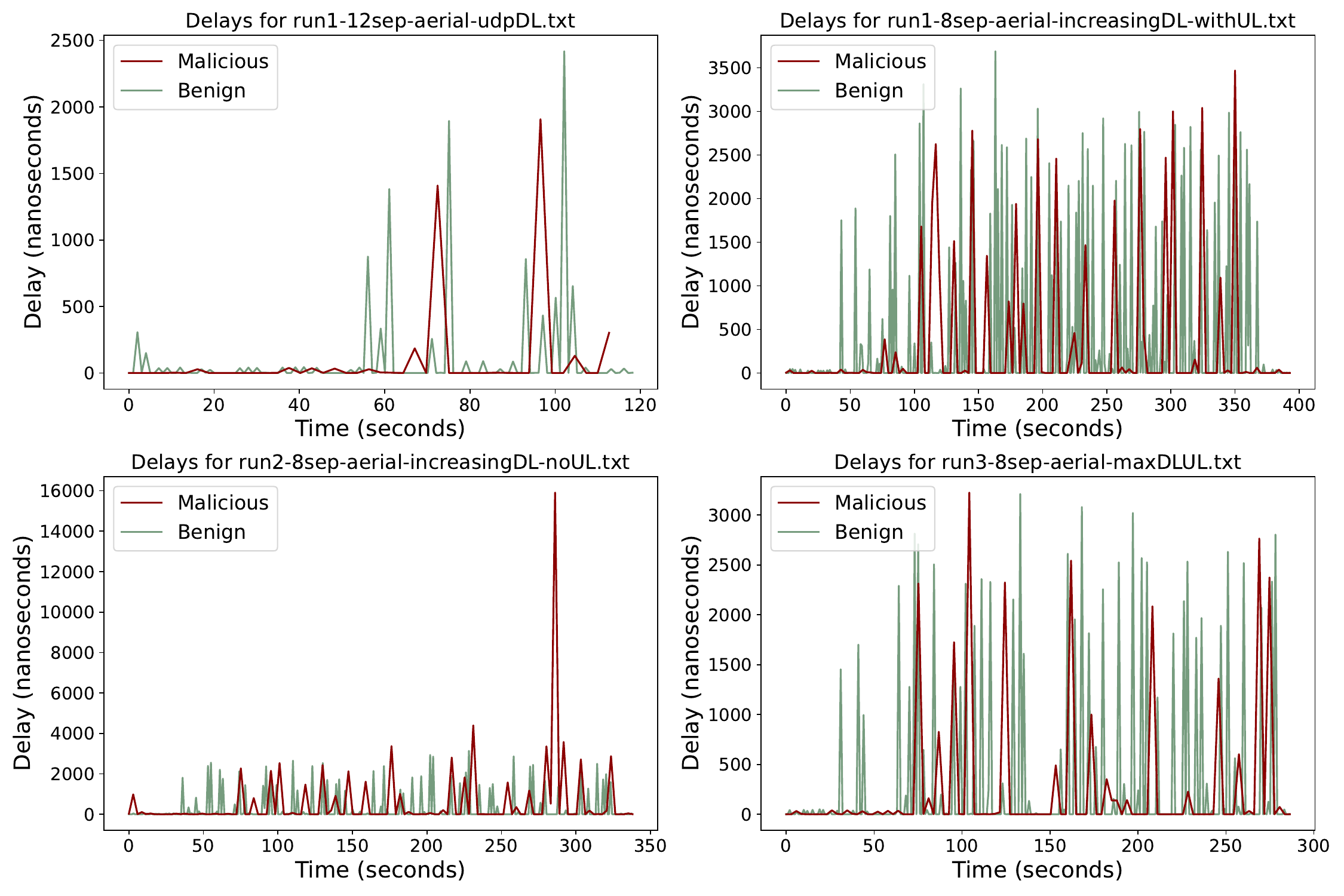}
    \caption{Impact of the 40/20 second spoofing attack. }
    \Description{A graph of the PTP reported delay during an attack.}
    \label{fig:Spoofing40/20}
\end{figure}

\begin{figure}[htb]
    \centering
    \includegraphics[width=.9\linewidth]{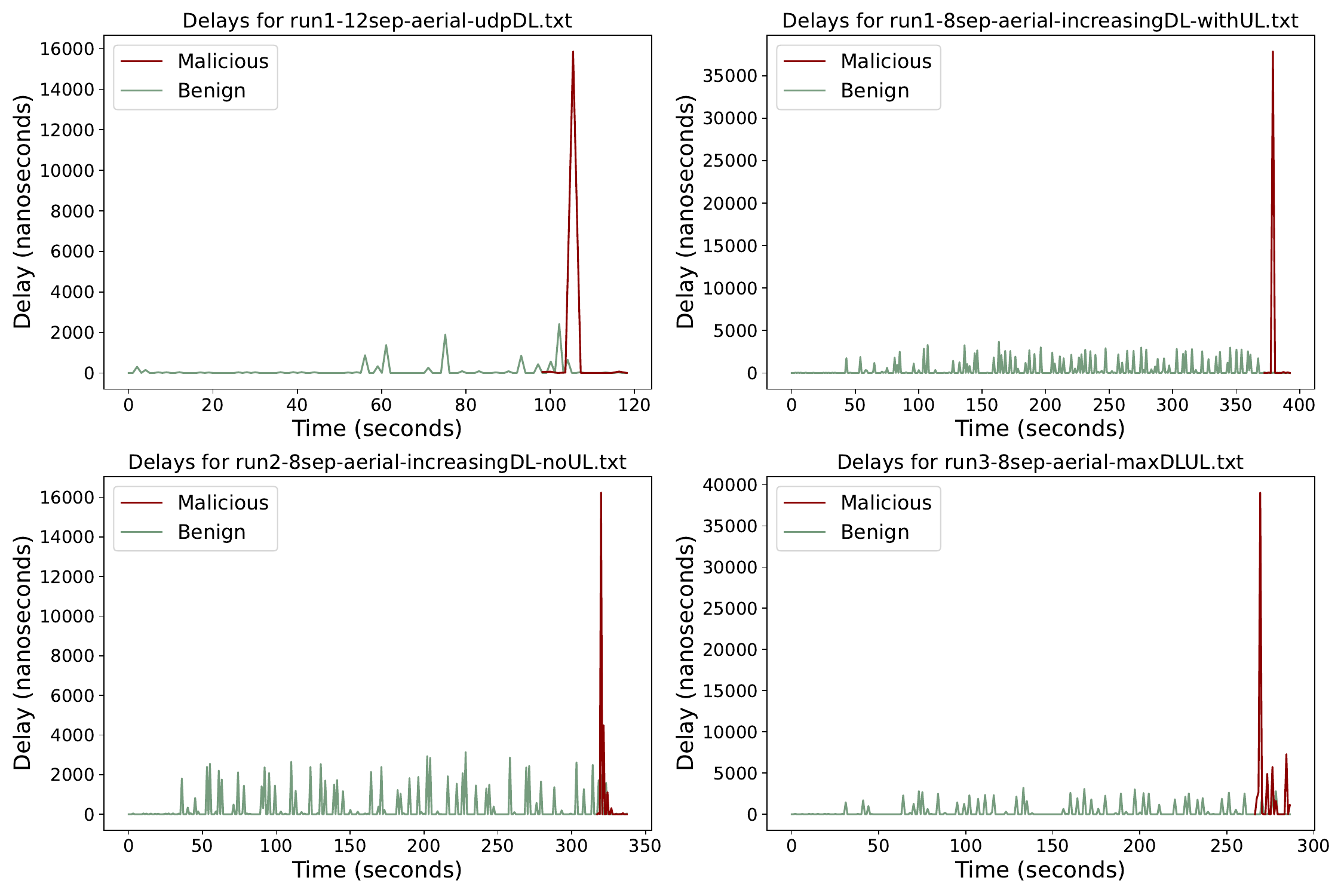}
    \caption{Impact of the continous spoofing attack. }
    \Description{A graph of the PTP reported delay during an attack.}
    \label{fig:Spoofing_cont}
\end{figure}

Figures \ref{fig:Spoofing30/30}, \ref{fig:Spoofing40/20}, and \ref{fig:Spoofing_cont} illustrate the impact of our spoofing attack. In Fig.~\ref{fig:Spoofing_cont}, the attack's impact is visible only after it ends because it stops \gls{ptp} synchronization messages. The longer the attack lasts, the greater the synchronization drift. After the attack, the large spike in delay shows the synchronization drift that occurred.

\begin{figure}[htb]
    \centering
    \includegraphics[width=.9\linewidth]{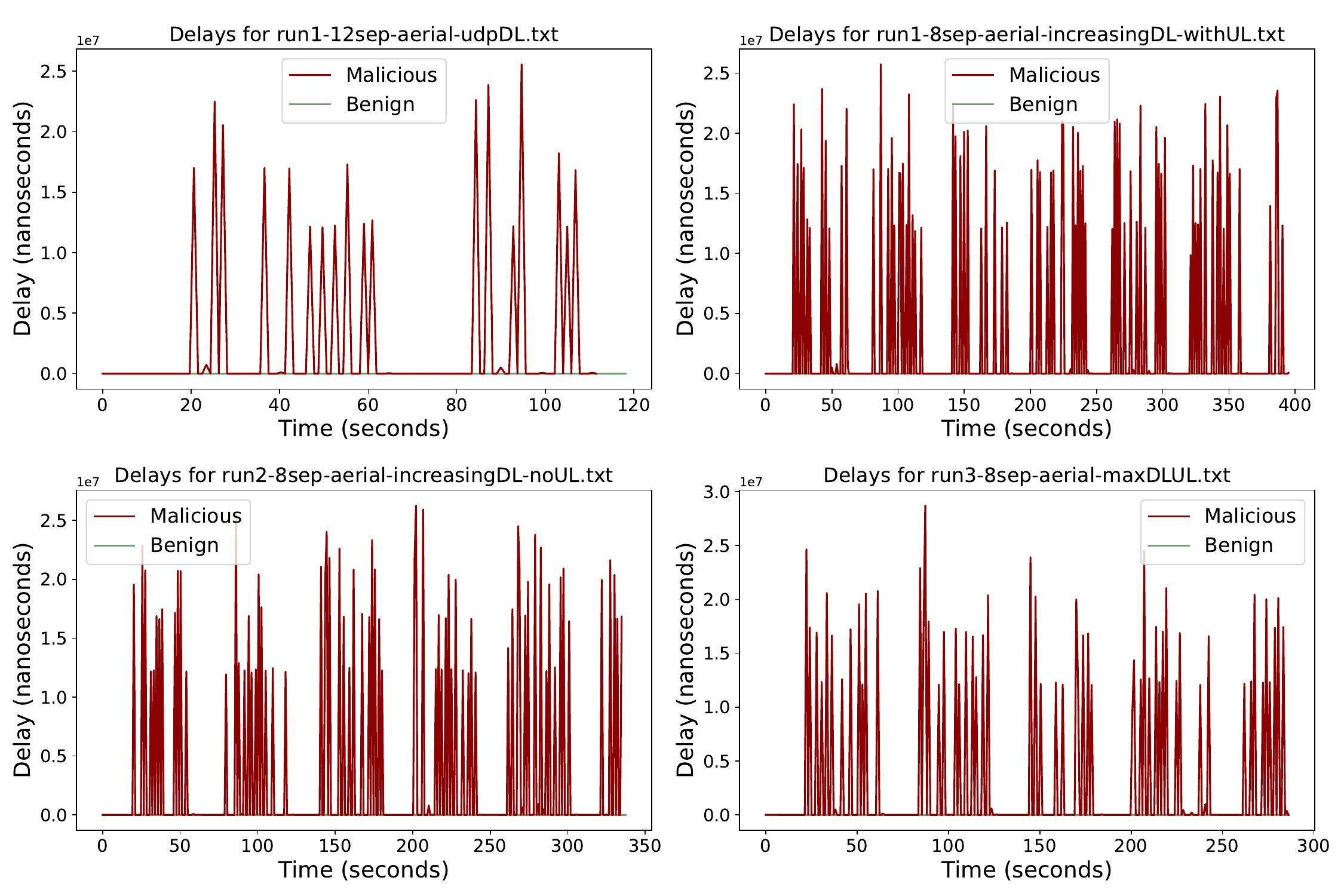}
    \caption{Impact of the 40/20 replay attack. }
    \Description{A graph of the PTP reported delay during an attack.}
    \label{fig:Replay40/20}
\end{figure}

\begin{figure}[htb]
    \centering
    \includegraphics[width=.9\linewidth]{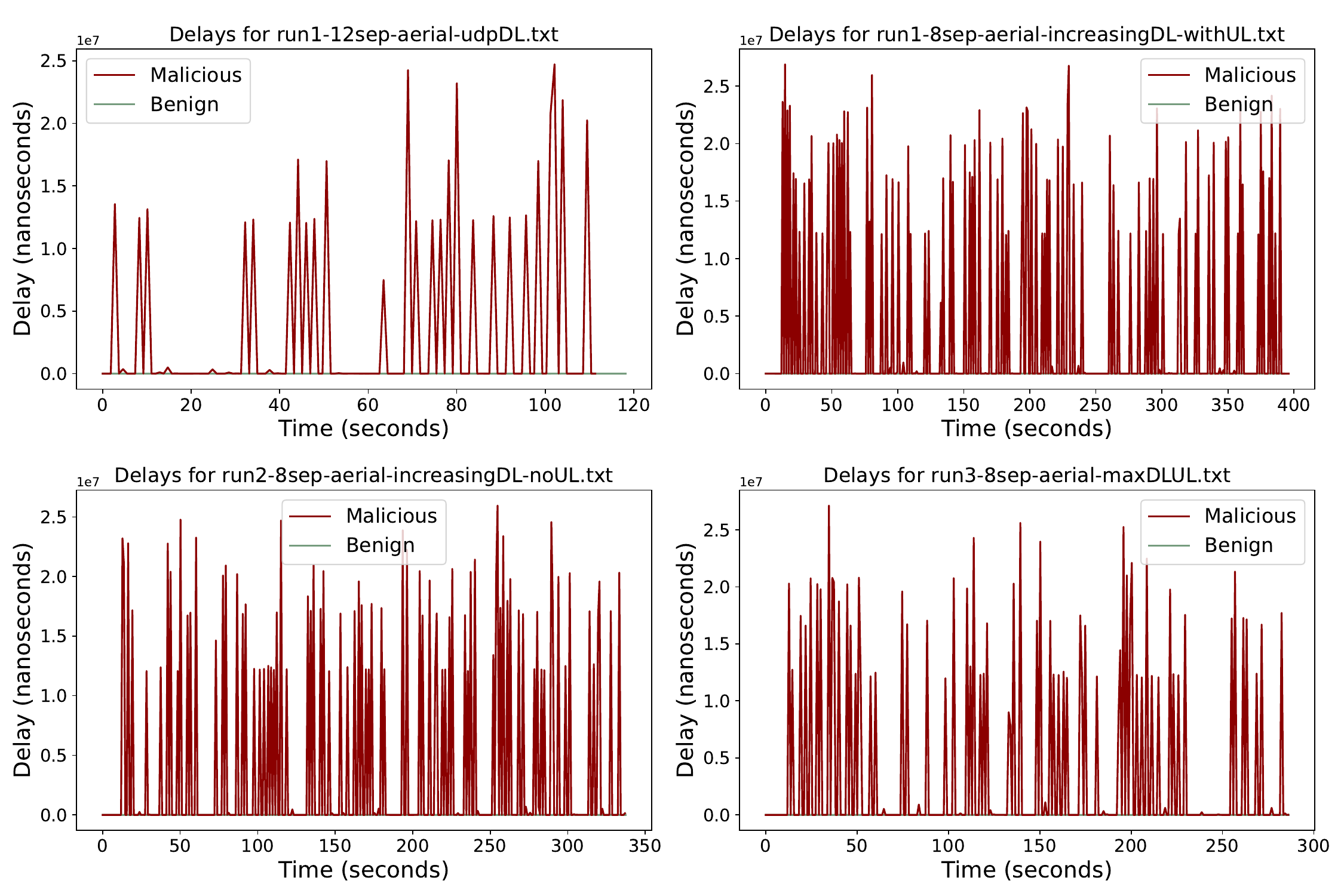}
    \caption{Impact of the 50/10 replay attack. }
    \Description{A graph of the PTP reported delay during an attack.}
    \label{fig:Replay50/10}
\end{figure}

\begin{figure}[htb]
    \centering
    \includegraphics[width=.9\linewidth]{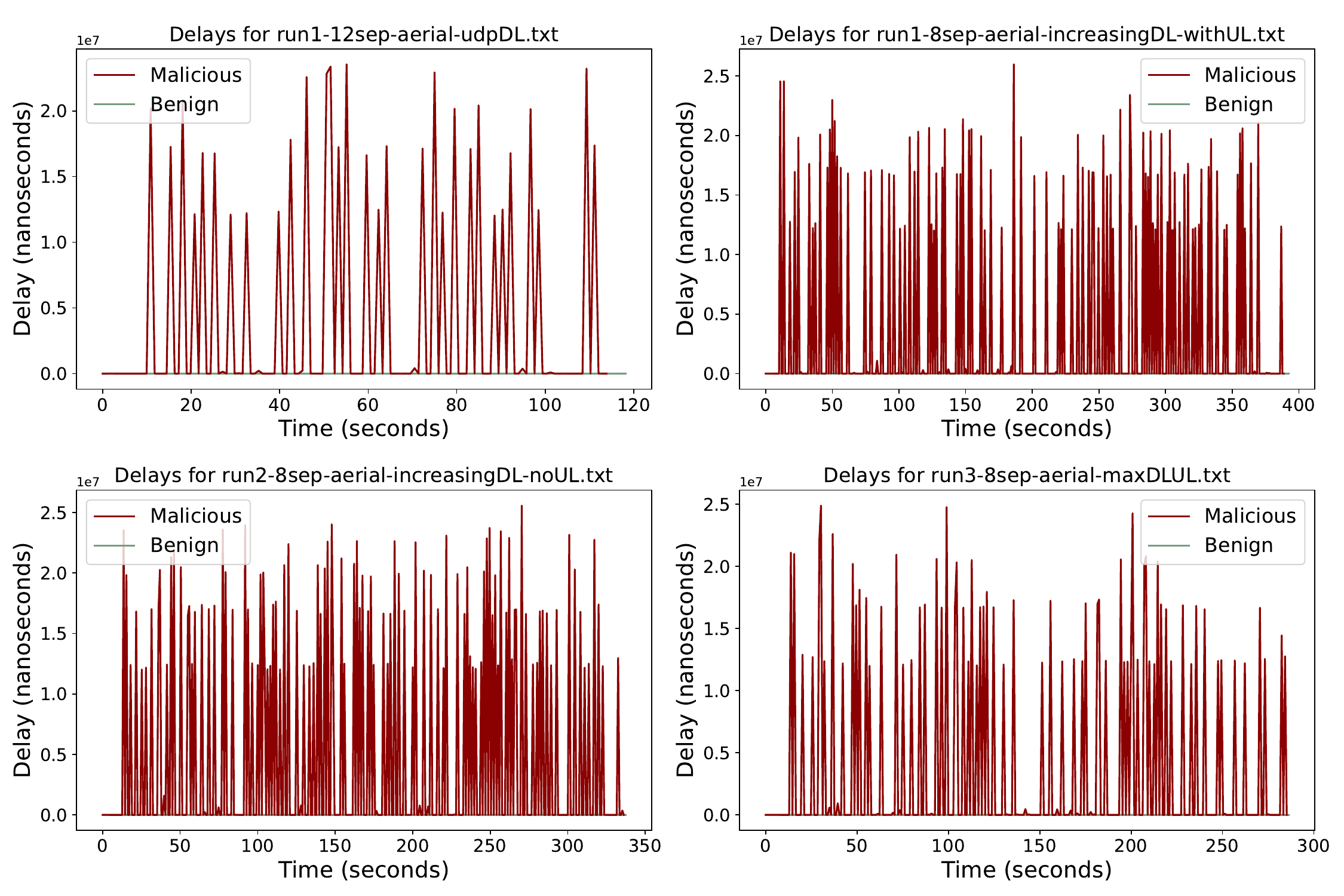}
    \caption{Impact of the continuous replay attack. }
    \label{fig:Replay_cont}
    \Description{A graph of the PTP reported delay during an attack.}
\end{figure}

Figures \ref{fig:Replay40/20}, \ref{fig:Replay50/10}, and \ref{fig:Replay_cont} show the impact of our replay attack. Unlike the announce attack, \gls{ptp} traffic continues during the replay attack, allowing delay calculation throughout. However, the reported delay is far higher than the actual network delay, severely impacting \gls{ptp} synchronization with reported delays over 8000 times greater than the actual observed delay.

\subsection{PLFS}\label{sec:plfs}

In the context of the \system framework, it is important to discuss \gls{plfs}. While \gls{fh} standards~\cite{OranWG11-secreqspec, etsiFh} allow for any \gls{plfs}, in practice, Synchronous Ethernet (SyncE) is commonly used. SyncE is a physical layer protocol~\cite{G.8262} that enables high-precision clock synchronization by using the edges in the Ethernet data signal to define the timing content of the signal. This protocol distributes a physical layer clock across a packet network, with each system recovering and forwarding the network timing through the distribution path. SyncE is primarily utilized to assist in frequency synchronization, ensuring that all network elements are aligned to a common frequency reference.

Our production-ready environment employs SyncE, providing robust frequency synchronization to support the timing requirements of \gls{ptp}. However, despite the presence of SyncE, our attack against \gls{ptp} was successful, demonstrating the potential vulnerability even with physical layer support. Our digital twin environment does not currently use SyncE, as our focus is on evaluating the security and resilience of the higher layer \gls{ptp} protocol. This approach allows us to isolate and test the specific impacts of attacks on \gls{ptp} synchronization without the additional complexity introduced by physical layer frequency support.

\subsection{ML Model Parameters}\label{a:model parameters}
In this section we give a detailed description of each model used.

\paragraph{Logistic Regression:} 
The Logistic Regression model from scikit-learn is a linear model for binary classification that estimates the probability of a binary outcome using the logistic function. The model is configured with the following parameters: \say{penalty=l2} for L2 regularization, \say{dual=False} to use the primal formulation for the optimization problem, \say{tol=1e-4} as the tolerance for stopping criteria, \say{C=1.0} as the inverse of regularization strength, \say{fit\_intercept=True} to add intercepts to the decision function, \say{solver=lbfgs} for optimization, \say{max\_iter=100} as the maximum number of iterations for the solver to converge.

\paragraph{k-Nearest Neighbors (k-NN):} 
The k-Nearest Neighbors model is a non-parametric method for classification that identifies the most common class among the \say{k} nearest neighbors in the feature space. The model is configured with the following default parameters: \say{n\_neighbors=5} as the number of neighbors to use for k-neighbors queries, \say{weights=uniform} to weight all points in each neighborhood equally, \say{algorithm=auto} to automatically select the best algorithm (\say{ball\_tree}, \say{kd\_tree}, or \say{brute}) based on the input data, \say{leaf\_size=30} as the leaf size for tree-based algorithms, and \say{p=2} as the power parameter for the Minkowski distance metric, equivalent to the Euclidean distance.

\paragraph{Decision Tree Classifier:} 
The Decision Tree model is a non-parametric supervised learning method that uses a series of decision rules inferred from data features. The model is configured with \say{criterion=gini} as the splitting criterion based on Gini impurity, \say{splitter=best} to select the best split, \say{max\_depth=None} to expand nodes until all leaves are pure or contain fewer than \say{min\_samples\_split} samples, \say{min\_samples\_split=2} as the minimum number of samples required to split an internal node, \say{min\_samples\_leaf=1} as the minimum number of samples required at a leaf node, \say{max\_features=None} to consider all features for the best split, and \say{random\_state=None} for randomization.

\paragraph{Gradient Boosting Classifier:} 
The Gradient Boosting Classifier is an ensemble learning method that builds multiple decision trees sequentially, where each tree corrects the errors of the previous one. The model is configured with \say{n\_estimators=100} as the number of boosting stages (trees), \say{learning\_rate=0.1} as the step size for weight updates, \say{max\_depth=3} as the maximum depth of each decision tree, and \say{verbose=1} to control the verbosity of the training output. Other parameters include \say{criterion=friedman\_mse} for measuring the quality of a split, \say{min\_samples\_split=2} for the minimum number of samples required to split an internal node, \say{min\_samples\_leaf=1} for the minimum number of samples required at a leaf node, \say{max\_features=None} to consider all features for the best split, and \say{random\_state=None} for randomization.

\paragraph{Naive Bayes Classifier:} 
The Gaussian Naive Bayes model is a probabilistic classifier that applies Bayes' theorem, assuming feature independence given the class. The model is configured with \say{priors=None} to use the class priors adjusted according to the data, and \say{var\_smoothing=1e-9} to set the portion of the largest variance of all features that is added to variances for calculation stability.

\paragraph{Random Forest Classifier:} 
The Random Forest Classifier is an ensemble learning method that combines multiple decision trees, trained on different parts of the data, for improved accuracy and reduced overfitting. The model is configured with \say{n\_estimators=100} as the number of trees in the forest, \say{criterion=gini} as the splitting criterion based on Gini impurity, \say{max\_depth=None} to expand nodes until all leaves are pure or contain fewer than \say{min\_samples\_split} samples, \say{min\_samples\_split=2} as the minimum number of samples required to split an internal node, \say{min\_samples\_leaf=1} as the minimum number of samples required at a leaf node, \say{max\_features=sqrt} setting the number of features to consider when looking for the best split to $\sqrt{6}$, and \say{random\_state=None} for randomization.

\paragraph{LSTM Classifier:} 
The LSTM model is designed with a single LSTM layer, followed by a fully connected (FC) layer. The LSTM layer is configured with \say{batch\_first=True}, allowing the input and output tensors to have the shape \say{(batch\_size, sequence\_length, input\_dim)}, which facilitates mini-batch training. The LSTM accepts a variable sequence length input ranging from 10 to 40 packets, enabling it to produce a regular, time-based output while processing a variable number of input packets. The hidden layer's dimensionality is defined by the \say{hidden\_dim} parameter, and the model outputs a binary classification for each input packet.

The output layer is a fully connected layer that maps the hidden states from the LSTM layer to a single output node for each input packet. The model uses a sigmoid activation function at the final layer to ensure the output is a probability value, making it suitable for binary classification tasks.

During training, the slice length is randomly selected from the range [10, 40] for each epoch. A custom Weighted Binary Cross-Entropy Loss function is employed to handle class imbalance and unequal error costs. For the LSTM model, the loss weight for false positives (\gls{fp}) is kept constant at 1, while the weight for false negatives (\gls{fn}) is varied between 1 and 10, achieving maximum recall with a value of 7. The Adam optimizer is used with an initial learning rate of 0.001, and a StepLR scheduler reduces the learning rate by a factor of 0.1 every 10 epochs to help the model converge more effectively. Early stopping is enabled by monitoring validation loss with a patience parameter of 20 epochs.

\paragraph{CNN Model:} 
The Convolutional Neural Network model is designed with a 2D convolutional architecture. The model processes input sequences of fixed length $S=32$ with features $F=6$, formatted as a single-channel image of dimensions $S \times F$. The model consists of two convolutional layers.

The first convolutional layer applies 16 filters, each with a kernel size of $3 \times 3$, using a stride of 1 and padding of 1 to preserve the input's spatial dimensions. This is followed by a ReLU activation function and a max pooling operation with a $2 \times 2$ kernel, which reduces the spatial dimensions by a factor of 2. The second convolutional layer uses 32 filters, again with a $3 \times 3$ kernel, and follows the same ReLU activation and max pooling steps. After the second layer, the feature map dimensions are further reduced.

The output from the convolutional layers is flattened to create a vector input for the fully connected (FC) layers. The first FC layer contains 64 neurons with a ReLU activation function, and the final FC layer outputs a single value, which is passed through a sigmoid activation function to produce a binary classification. The model is trained to minimize a custom Weighted Binary Cross-Entropy Loss function to handle class imbalance and uneven costs of false positives and false negatives, although we kept both weights at the default setting of 1.

The Adam optimizer is employed with an initial learning rate of 0.001, and a StepLR scheduler reduces the learning rate by a factor of 0.1 every 10 epochs. Early stopping is enabled by monitoring validation loss with a patience parameter of 20 epochs.

\paragraph{Transformer Model:} 
The transformer-based model leverages the encoder portion of the transformer architecture, omitting the decoder, to process input sequences with fixed lengths of $S = 16$, $32$, or $40$ packets during training and evaluation. These lengths are chosen to balance the trade-off between the need to incorporate more historical data, which improves accuracy and recall, and the associated computational cost and memory requirements.

The encoder is composed of a stack of $N=2$ layers, each layer consisting of a multi-head self-attention mechanism with $n_{head}$ attention heads, followed by a feed-forward network. The input sequence has six features, and is normalized by a layer normalization step before being processed by the encoder layers. The encoder outputs a contextualized feature matrix of size $F \times S$, where $F$ is the number of features and $S$ is the sequence length. 

The encoded output is then flattened and passed through a fully connected linear layer with 256 neurons, followed by a ReLU activation function to introduce non-linearity. After this, a dropout layer with a dropout probability of 0.2 is applied to mitigate overfitting. The final output is generated by another fully connected layer with a single output neuron, followed by a sigmoid activation function, to produce a binary classification result for the entire input sequence.

We treated the number of attention heads ($n_{head} = 2, 3$) and the slice length ($S = 16, 32, 40$) as hyper-parameters and tuned them for optimal recall. The model is trained using the Adam optimizer with an initial learning rate of 0.001. A StepLR scheduler is used to reduce the learning rate by a factor of 0.1 every 10 epochs, facilitating convergence by fine-tuning the learning rate during training. Early stopping is enabled by monitoring the validation loss with a patience parameter of 20 epochs. We use the same custom Weighted Binary Cross-Entropy Loss function with both weights kept constant at 1.

\end{document}